\documentclass{aa}  

\usepackage{footnote}
\makesavenoteenv{figure*}

\newcommand{\dd}{\mathrm{d}}
\usepackage{graphicx}
\usepackage{subcaption}

\usepackage{txfonts}
\usepackage{hyperref}
\hypersetup{
colorlinks=true,
linkcolor=blue,
filecolor=blue,
citecolor = blue,
urlcolor=cyan,
}

\begin{document} 

\title{The signature of major mergers on the hydrostatic mass bias of galaxy clusters}
  
   \titlerunning{Signature of major mergers on the hydrostatic mass bias}

   \author{Isac Barranco-Llorca
          \inst{1}
          \and
           David Vallés-Pérez\inst{2,1}
           \and
           Susana Planelles \inst{1,3}
           \and
           Vicent Quilis \inst{1,3}
          }

   \institute{Departament d'Astronomia i Astrofísica, Universitat de València, 46100 Burjassot, Spain\\
              \email{Isac.barranco@uv.es}
         \and
             Dipartimento di Fisica e Astronomia, Università di Bologna, Via Gobetti 93/2, IT-40129 Bologna, Italy
         \and 
            Observatori Astronòmic, Universitat de València, 46980 Paterna (València), Spain
            }

   \date{Received X XX, XXXX. Accepted X XX, XXXX}

    \abstract
    {While the masses and abundances of galaxy clusters are key observables for Cosmology, mass determinations based on intra-cluster medium observations often rely on the hydrostatic equilibrium  assumption, introducing a systematic bias, which is expected to be more noticeable during strong assembly episodes such as major mergers.}
    {We aim to study in detail how mergers shape the variability in the hydrostatic mass bias values through the evolutionary history of clusters, identify the primary mechanisms driving such evolution, and assess its potential dependence on dynamical state and merging history.}
    {Using a moderate-sized AMR Eulerian+$N$-Body cosmological simulation, we identify a sample of galaxy cluster mergers in the redshift interval $1.5 \leq z \leq 0$. We compare true and hydrostatic masses within the virial volume, the latter derived from gas density and temperature radial profiles. The evolution is assessed in relation to the merging history, extracted from halo merger trees.}
    {At $z=0$, the hydrostatic mass bias shows a mild correlation with dynamical state. During major mergers, the bias follows a characteristic trend: a sharp negative dip around the merger time, a transient positive peak, and a gradual return to pre-merger levels. This behaviour is primarily driven by morphological and dynamical reconfigurations of the gas density within the ICM, while thermodynamical processes play a secondary role. The pattern shows no strong dependence on secondary parameters, such as mass ratio or impact parameter, but it can be fitted to a simple time-dependent functional form. This trend is present at radii  $r\le R_{\mathrm{vir}}$, although with reduced amplitude and shorter timescales as the radius decreases.}
    {Hydrostatic mass bias is closely linked, albeit in a non-trivial way, with the merging history of galaxy clusters. 
    We find that the bias values are weakly correlated with the dynamical state of clusters. Nevertheless, our results give a robust estimation of the hydrostatic mass bias values in the pre-merger, merging, and post-merger phases. These findings highlight the importance of delving deeper into the observational assessment of cluster assembly state in order to improve mass estimations for cosmological analyses.}

  
   \maketitle
%

\section{Introduction}

Galaxy clusters, situated at the cusp of hierarchical structure formation, conform the biggest gravitationally bound objects that have decoupled from cosmic expansion. These structures set the largest known scales at which baryons --and, thus, hydrodynamics-- significantly influence a Universe dominated by dark matter (DM) and dark energy. Therefore, galaxy clusters offer an exceptional environment in which Astrophysics and Cosmology meet each other, serving as cosmological-scale laboratories where to study the formation and evolution of cosmic structures \citep[e.g.,][]{Kravtsov_2012, Planelles_2014, Walker_2019}.

Within this context, cluster mass is perhaps their most fundamental quantity. On the one hand, it is closely connected to their internal structure \citep{Kaiser_1986, Lau_2015}. On the other hand, the abundance of clusters as a function of mass is a powerful probe of the cosmic matter density fraction ($\Omega_m$) and the amplitude of primordial density fluctuations ($\sigma_8$; e.g. \citealp{Allen_2011}, for a review). In this regard, the accuracy of mass estimates directly limits the precision of cosmological parameter constraints and the ability to test theoretical models \citep{Mantz_2015, Zubeldia_2019}.

Since cluster mass is not directly observable, it must be inferred from other information. Various indirect approaches are employed to estimate cluster mass within surveys, depending on the observable of choice and the specific assumptions considered. These include using gravitational weak lensing, kinematic analysis of member galaxies, X-ray continuum emission and/or spectral lines, and the thermal Sunyaev-Zel'dovich (SZ) effect, among others, to calibrate the observables and infer masses across broader samples (we refer the reader to \citealp{Pratt_2019} for an in-depth review on observational mass estimation methods). Even the weak-lensing method, which might be considered the more direct, since it is a direct probe of the gravitational potential, is still subject to uncertainties due to mass modelling and projection effects  \citep[e.g.,][]{Hoekstra_2013, Umetsu_2020}. It is due to this fact that each of the different mass estimation methods may introduce various biases, on top of their considerable intrinsic scatters, making it crucial to identify and characterise these systematic effects in order to prevent them to propagate to subsequent cosmological or astrophysical results.

Observations of the  intra-cluster medium (ICM) in either X-ray or microwaves (SZ) can be used to infer its temperature, electron density and/or thermal pressure three-dimensional radial profiles, using suitable deprojection methods and assuming spherical symmetry \citep[e.g.,][]{Ettori_2013}. From these profiles, the total mass can be estimated under the hydrostatic equilibrium (HE) assumption \cite[e.g.,][]{Sarazin_1986,Markevitch_1996,Vikhlinin_2006,Pratt_2019}. Such conditions are usually grounded on the --ideal-- expectation that an isolated cluster will gradually settle into an equilibrium state over several dynamical times \citep{Poole_2006}. However, galaxy clusters are dynamically young objects, embedded in a highly anisotropic cosmological environment and subject to continuous accretion of matter and more violent merging events, even by $z \simeq 0$.
When additionally considering the effect of baryons, the presence of both, bulk and turbulent, motions, or thermal and kinematic feedback from supernovae and active galactic nuclei, raise the complexity of the system to levels where the former assumptions are likely to be too simplistic and not representative of the real ICM \citep[e.g.,][]{Nagai_2007_a,Nelson_2012}.
In this sense, the path to a relaxed state is far from being trivial, and may not even be achieved (see, e.g., \citealp{Poole_2006, ZuHone_2011}, for idealised cluster merger simulations; or \citealp{Nelson_2014, Bennett_2022}, more particularly on the topic of HE).

In consequence, neither the sphericity nor the HE assumptions are regularly met, thus introducing deviations from the true mass. More in particular, the presence of nearly-isotropic turbulent motions, that can formally resemble a pressure term in the motion equation, implies that less thermal pressure is required for resisting gravitational collapse than predicted by the HE approximation. Hence, neglecting this contribution  tends to underestimate the true mass.  

Formally, the hydrostatic mass bias, $b$, can be defined\footnote{Different definitions for the hydrostatic mass bias can be found throughout the literature, differing from the one we use in this paper as $b' = -b$ or $b' = 1-b$.} as the normalised difference between the total mass, $M_\text{true}$ (true gravitating total mass, obtained directly from  simulations as the sum of baryons and DM),  and the hydrostatic mass, $M_\text{HE}$ (derived from  density and temperature profiles):

\begin{center}
\begin{equation}
    b=\frac{M_\text{HE}- M_\text{true}}{M_\text{true}}.
    \label{Mass bias formula}
\end{equation}
\end{center}

In order to characterise these offsets and find proper corrections applicable to observational data, over the last two decades, many studies have looked at the deviations from the HE condition in hydrodynamical simulations,  either by directly computing HE mass estimates and comparing to true mass profiles \citep{Nagai_2007, Biffi_2016, 2020Angelinelli, Bennett_2022} or by estimating the non-thermal contributions to pressure and comparing them to the purely thermal pressure  \citep[e.g.,][]{Lau_2013, Vazza_2018, 2020Angelinelli, Groth_2025, Lebeau_2025}.
These latter studies show how, throughout the whole radial extent of the cluster, there are significant ($\sim 0-40 \%$) contributions to the pressure support from bulk and small-scale motions, increasingly relevant at larger radii.

Notably, \citet{Biffi_2016} found that cool-core and non–cool-core clusters exhibit mild differences in their hydrostatic mass bias profiles, albeit these differences are restricted to the innermost regions ($r \lesssim R_{2500c}$). Conversely, the distinction between regular and disturbed systems appears less significant. Moreover, they identified a strong correlation between the hydrostatic mass bias and deviations from HE. 
\citet{2020Angelinelli} studied the relation of turbulence, bulk motions and radial accelerations with mass bias, reporting that the relation turns out to be far from simple, and the trends are obscured by the object-to-object variance.

Complementing these findings, \citet{Groth_2025} explored the contribution of turbulent motions to the overall pressure support using varying simulation and analysis methods. While their results suggest a connection to dynamical state, the intrinsic scatter, together with the dependence on simulation and analysis choices, again confirm that this connection is far from trivial.

Owing to the complexity of ICM physics, the sources of mass bias vary dramatically along the evolution of a cluster, and from cluster to cluster. However, mergers produce perhaps the most remarkable dynamical and morphological perturbations, since a major merger can significantly disturb the host halo with an energy input up to  $\sim GM^2/R \sim 10^{64} \, \mathrm{erg}$. Part of the bulk kinetic energy involved in the merger is dissipated through merger shocks, heating up the ICM \citep[e.g.,][]{Vazza_2009, Planelles_2013}. Simultaneously, merger-induced motions cascade down to smaller scales and increase the degree of turbulent motions \citep[e.g.,][]{Vazza_2012, Valles-Perez_2021}. Both of these phenomena increase the pressure support on their own timescales, which additionally may depend on the merger configuration, ICM microphysics, etc.

This scenario highlights the importance of studying galaxy cluster mergers within a realistic cosmological context \citep[e.g.,][]{Planelles_2009}, where not only merger events but also the natural evolution of the cluster -- including smooth mass accretion, interaction with the cosmic filamentary network, and progression towards a dynamically relaxed state -- can influence the mass bias in distinct and complex ways.

In this work,  we investigate the temporal evolution of the mass bias at the virial radius along majors mergers. 
We do so by extracting a sample of 74 well-isolated galaxy cluster major merger events from a high-resolution cosmological simulation. 
Unlike previous works that focused on individual clusters, or bias measurements at a single redshift, this study is specifically set to explore how the hydrostatic mass bias evolves over time from the initial infall to the post-merger relaxation phase. 
Moreover, we also look at the overall trends of this bias with dynamical state and accretion history. 
These features offer indirect insight into the processes that influence mass bias, although further work is needed to isolate the specific physical mechanisms involved. In this way, our study complements and extends the findings of earlier works, providing a statistical study on the temporal evolution of the hydrostatic mass bias in dynamically active periods.

This paper is organised as follows. In Sect.~\ref{methods}, we present the details of the simulation we have studied and the methods we have used to treat and analyse the data. In Sect.~\ref{results}, we explore the hydrostatic mass bias and its evolution through major merger events. Finally, Sect.~\ref{conclusions}, contains a summary of our findings. We also have included Appendix \ref{appendix overlaping mergers}, where we analyse a representative cluster exhibiting a rather complex merging history.

\section{Methods}
\label{methods}

In this section we present the numerical methods and the data treatment we have employed in order to study the simulation output. In Sect. \ref{simulation studyed}, we introduce the details of the cosmological simulation. To continue, in Sect. \ref{catalogues and mtrees}, we present the algorithms to identify the galaxy clusters and to build  their merger trees, culminating in the selection of a suitable sample of mergers for the statistical study. Then, we describe the treatment and calculation of the mass profiles in Sect. \ref{mass profiles}, and in Sect. \ref{dyn state} we explain how we characterise the dynamical state of the clusters.

\subsection{Simulation details} 
\label{simulation studyed}

The simulation analysed in this work, which was already used for previous studies of ICM physics and galaxy cluster assembly (see \citealp{Valles-Perez_2024, Valles-Perez_2025} for further details), has been performed with the cosmological code \texttt{MASCLET} \citep{Quilis_2004, Masclet_B}. The code integrates (magneto-) hydrodynamics and $N$-body techniques to evolve both gaseous and DM components. The gaseous (collisional) component is treated with high-resolution shock-capturing schemes, while the DM (collisionless) component is evolved through a multilevel Particle-Mesh \citep[PM;][]{Hockney_Eastwood_1988} $N$-body algorithm. A self-consistent gravity solver ensures the coupled evolution of both components. Additionally, \texttt{MASCLET} incorporates an adaptive mesh refinement (AMR) scheme to enhance spatial and temporal resolution.

The simulation assumes a flat $\Lambda$CDM cosmology with the following parameters: total matter density ${\Omega_{m}=0.31}$, baryon density ${\Omega_b = 0.048}$, dark energy density ${\Omega_{\Lambda}=1-\Omega_{m} = 0.69}$, and a dimensionless Hubble parameter \mbox{$h \equiv H_0 / (100 \; \mathrm{km \; s^{-1} \; \mathrm{Mpc}^{-1}}) = 0.678$}.
 
Specifically, the computational volume consists on a periodic, cubic domain of comoving side length $L=147.5$ Mpc, discretized  with a grid of $256^3$ uniform cells at the coarse level of resolution ($l=0$).
 The initial conditions are generated at redshift $z_\mathrm{ini}=100$ from a realization of a Gaussian random field with spectral index $n_s=0.96$ and amplitude $\sigma_8=0.82$, using the CDM transfer function of \citet{Eisenstein_1998}. These values are consistent with the latest results from the Planck Collaboration \citep{Planck_2020}.

The simulation is first evolved at low resolution with equal-mass DM particles to the present time, from which regions meeting specific refinement criteria are selected for further refinement. This initiates three nested AMR levels ($l=1,2,3$), with DM particles that are 8, 64, and 512 times less massive, respectively, than those used in the base level ($l=0$), already at $z_\mathrm{ini}$. As the system evolves, the local baryonic and DM densities determine where finer grids are introduced, reaching a maximum refinement level at $l=6$. Each level refines spatial resolution by a factor of two (${\Delta x_{l+1}/\Delta x_l=1/2}$), balancing numerical stability with the gain in resolution. This allows a peak spatial resolution of ${\Delta x_{l=6} = 9\,\mathrm{kpc}}$ at $z=0$.

Regarding the implementation of DM, the simulation uses four different species of particles, achieving a peak mass resolution of about $\sim 1.5\times10^7\,M_{\odot}$ equivalent to using $2048^3$ particles throughout the computational domain.

In addition to gravity and hydrodynamics, the simulation incorporates several physical processes: cooling (free-free, inverse Compton and atomic and molecular cooling for primordial gas) and heating mechanisms \citep[such as UV background radiation as in][]{Haardt_Madau_1995}.
However, the simulation does not account for star formation nor any additional feedback mechanisms.

\subsection{Cluster catalogues and merger trees}
\label{catalogues and mtrees}

The simulation is post-processed using the publicly available\footnote{https://github.com/dvallesp/ASOHF.} \textit{Adaptive Spherical Overdensity Halo Finder} \texttt{ASOHF} \citep[see][for more details]{asohf1, Knebe_2011, asohf2}, which is employed to generate a catalogue of dark matter halos for each snapshot.

Halos across different snapshots are linked using the supplementary script \texttt{mtree.py} from the \texttt{ASOHF} package, which identifies all halos from preceding snapshots contributing to the halo mass at a given time. This allows us to track the assembly history of DM structures across time and build a consistent evolutionary picture for each halo.

In this study, we focus on a sample of isolated and well-defined major merger events, extracted from the simulation in the $z \in [0, 1.5]$ interval.

The selected events involve host halos with virial masses above ${4\times10^{13} \, M_{\odot}}$ in the pre-merger snapshot and include both galaxy clusters and massive groups, allowing us to explore merger-driven effects across a broad mass range. To identify major mergers, we adopt a mass ratio threshold of ${q \equiv M_{\text{infall}}/M_{\text{host}} > 0.2}$, slightly below the conventional 1$:$2 or 1$:$3 cuts \citep{Planelles_2009, Behroozi_2015}, but high enough to ensure a strong dynamical impact on the ICM. This threshold is consistent with previous works  \citep[e.g.,][]{Nelson_2012}, while enabling us to probe a wider diversity of merger dynamics. Thus, the cluster catalogue is formed by 427 objects that have suffered, at least, one major merger event in their lifetime and have met the minimum mass criterion.

To assess the effects of the major merger on the ICM properties with minimal contamination from other major mergers, it is necessary for the event to be temporally isolated. In order to do so, we compute the virial timescale (also referred to as the dynamical timescale) at the merger epoch as ${\tau_\text{vir}(z)= 1/\sqrt{G\Delta_{\mathrm{vir}}(z) \rho_c(z)}}$, where ${\rho_c(z)=3H^2(z)/(8\pi G)}$ is the critical density of the Universe at a given z, with ${H(z)=H_0\sqrt{\Omega_m(1 + z)^3 + \Omega_{\Lambda}}}$ the Hubble parameter, and ${\Delta_{\mathrm{vir}}(z) = 18\pi^2 + 82x - 39x^2}$, the virial overdensity with respect to the critical density of the Universe, where  ${x = \Omega_{m,0}(1+z)^3/(H(z)/H_0)^2 -1}$  \citep{Bryan_1998}.
Once we have computed the virial timescale, we require the merger to be isolated -- that is, be the single major merger taking place in ${t \in [t_\mathrm{merger}-\tau_\mathrm{vir}/2,t_\mathrm{merger}+\tau_\mathrm{vir}]}$ -- and also not to overlap with the previous merger dynamical time interval (in case there is one before). 

Moreover, to properly consider the redshift interval of interest, we exclude from the analysis those mergers that occur less than 3 snapshots (around $1 \, \mathrm{Gyr}$) from the time interval limits. In this way, we avoid considering mergers that may be contaminated by other major events taking place outside of the considered time domain. On the other hand, using the halo positions and their centre-of-mass velocities in the snapshots immediately prior and after the merger, and assuming a rectilinear trajectory derived from the relative motion of the two halos, we make a tentative prediction of the merger time (as the epoch when the centre of the infalling halo crosses the host virial radius) and of the impact parameter.

After this whole process, a sample of 74 properly isolated major mergers is obtained. 
It is important to note that our analysis is not limited to mergers involving clusters that belong exclusively to the main branch of a particular merger tree, that is, those that reach $z=0$. Instead, we also consider mergers between clusters whose evolutionary paths are intercepted at lower redshifts because of their subsequent incorporation into a main branch.

\subsection{Cluster hydrostatic mass profiles}
\label{mass profiles}

According to the HE prescription, gas in galaxy clusters can be considered in  equilibrium between thermal pressure  and gravitational forces, described by the Euler equation:
\begin{equation}
    \frac{\dd \textbf{v}}{\dd t} = -\nabla \Phi -\frac{1}{\rho} \nabla P \overset{\mathrm{HE}}{\approx} \mathbf{0}, 
    \label{euler eq}
\end{equation}
where $\rho$, $\textbf{v}$, $P$ and $\Phi$ stand, respectively, for the gas density, velocity field, thermal pressure and gravitational potential.

Thus, considering the equation of state for an ideal gas with only thermal pressure and constant mean molecular weight, $\mu$, the Poisson equation for the gravitational potential, $\Phi$, and further assuming spherical symmetry, the enclosed hydrostatic mass within a certain radius $r$ can be estimated using its thermodynamic observables: 

\begin{equation}
    M_\text{HE}(<r)= -\frac{k_B T(r)r}{\mu G m_p}\left(\frac{\dd\log \rho(r)}{\dd \log  r} + \frac{\dd \log  T(r)}{\dd \log  r}\right), 
    \label{mHE formula}
\end{equation}
where $k_B$ is the Boltzmann constant, $m_p$ is the proton mass and $T(r)$ and $\rho(r)$ are the temperature and density radial profiles of the gas, centred on the  overdensity peak of the cluster. These profiles are not obtained through traditional binning but via uniform angular sampling and interpolation on a fixed radial grid, which minimises the impact of substructures and ensures smooth, well-resolved profiles.

In order to get a mass estimation using Eq.~\ref{mHE formula}, for each of the clusters in our final sample we compute spherically averaged radial profiles for the gas density, the temperature, and the enclosed total mass from the halo centre out to a radius of $4R_{\mathrm{vir}}$. The derivatives in this equation are computed using the central difference method over the radial profiles smoothed with a \citet{Savitzky_1964} filter.

Let us note that Eq.~\ref{mHE formula} may allow negative mass estimates, especially in the innermost radii of halos,  
where density and temperature (or equivalently pressure) inhomogeneities may produce positive radial gradients, or in the presence of massive substructures.

Needless to say that, if any of the assumptions underpinning the model are not  satisfied,  the mass estimation obtained by means of such a model  will be over or underestimated, i.e., a difference between the HE mass profiles and the real mass profiles will arise, whose exploration is precisely the focus of this work.

\subsection{Dynamical state of clusters}
\label{dyn state}

As explained above, galaxy clusters are typically described as spherical systems, with the DM component in virial equilibrium and the ICM in HE. However, it has been demonstrated that this idealized description does not accurately capture the complexity of their actual structure and dynamics.
The morphological features and dynamical state of galaxy clusters are closely related to their assembly history, showing a close relation between merging,  strong accretion events, and the presence of substructures. Such merging events or any intense accretion period can trigger morphological and dynamical disturbances that, eventually, will be smoothed and fade away once the cluster moves towards equilibrium \citep[see, e.g.,][]{Poole_2006}.

Since dynamically relaxed and disturbed clusters exhibit significantly different characteristics, assessing their dynamical state is a crucial step when studying the HE condition and, therefore, its effects on the analysis of the mass bias. 
A wide range of physical parameters can serve as tracers of a cluster’s dynamical state, enabling different classification strategies to distinguish between different dynamical regimes.

In this work, we assess the dynamical state of galaxy clusters using a combination of several properties of their DM halo following the parametrization given in \cite{2025DVP}. The physical indicators, largely introduced in \cite{Dynamical_indicator_2023}, employed in this analysis to establish the relaxation degree of a galaxy clusters are: centre offset, virial ratio, mean radial velocity, ellipticity, fraction of mass in substructures and sparsity.

According to this method, in order to assess the degree of relaxedness of a cluster, we compute the relaxedness parameter $\chi$, which integrates the former six commonly-used individual dynamical state indicators via a redshift-dependent weighted combination. This combination is specially calibrated to correlate the dynamical state of galaxy clusters with merger events.

While the latter parameters intuitively measure the degree of disturbance of a cluster, 
the combined indicator $\chi$ points in the opposite direction, since
it measures the level of relaxedness.

\section{Results}
\label{results}

Firstly, in Sect.~\ref{mock test}, we motivate the subsequent study with simulations by presenting the results of a toy-model merger. Hereafter, in Sect. \ref{relax state sec}, we present the correlation of dynamical state with hydrostatic mass bias at $R_\mathrm{vir}$ for the sample of clusters at ${z=0}$. Next, in Sect. \ref{mass bias evolution} we continue with an exploration of the large variability of the mass bias during the lifetime of galaxy clusters. In Sect. \ref{bias in merger} we study the behaviour of the mass bias during major mergers, taking a closer look to the general trend on the whole sample in Sect. \ref{bias trend}, proposing an effective functional fit in Sect. \ref{merger trend fit} and exploring its underlying mechanisms in Sect. \ref{bias trend causes}.

\subsection{A toy model for hydrostatic mass bias evolution during mergers}\label{mock test}

The motivation for this work is rooted in the need to better understand how merger events influence the hydrostatic mass estimates of galaxy clusters. 

As this could be a complex scenario involving many physical ingredients, we decided to start a extremely simplified zero order approach  so as to investigate  idealized mergers of two objects sketching clusters. The aim would be to investigate whether the only variation on the orbital parameters of the infalling objects could imprint a characteristic pattern in the the behaviour of the hydrostatic mass bias under completely controlled and reproducible conditions.

In this mock merging scenario, both clusters are modelled as spherically symmetric systems, with static gas and DM density distributions drawn from the radial profiles of relaxed cluster from our sample which has had a constant mass bias of $\sim 20 \%$ for several Gyrs. The infalling halo, which is a rescaled version of the host, is treated as cold, lacking any temperature profile. In this sense, if the mass ratio is set to ${q \equiv M_\mathrm{vir, infall} / M_\mathrm{vir, host} < 1}$, we rescale the radius of the halo as $R_\mathrm{vir,infall} = q^{1/3} R_\mathrm{vir,host}$.
Finally, to mimic a major merger event, we model the trajectory of the infalling halo by modifying its position according to a Keplerian orbit, given an  initial position $\mathbf{r_\mathrm{ini}}$, impact parameter $p$, and initial velocity, ${v_\text{ini}=\sqrt{G M_\text{host}(1+q)/r_\text{ini}}}$.

\begin{figure}
  \centering
  \includegraphics[width=\hsize]{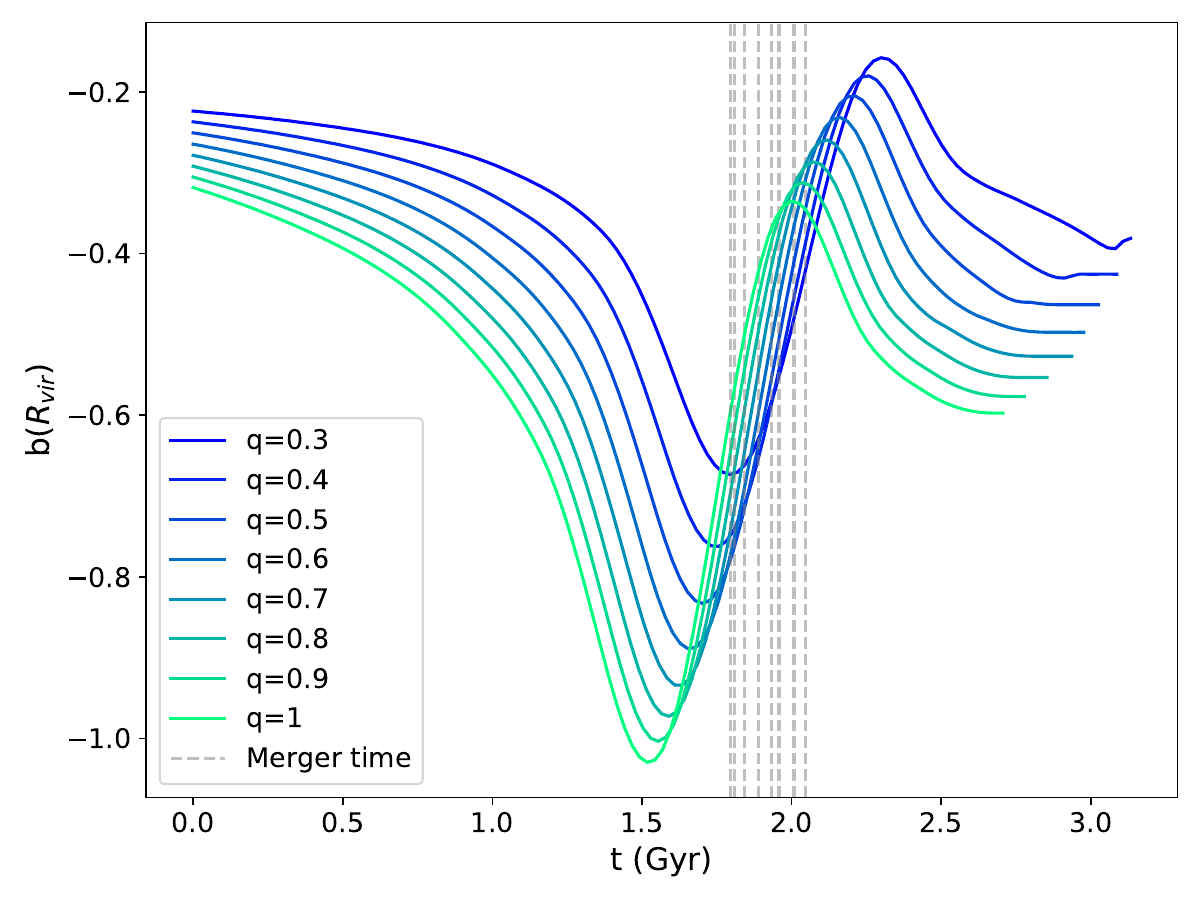}

  \centering
  \includegraphics[width=\hsize]{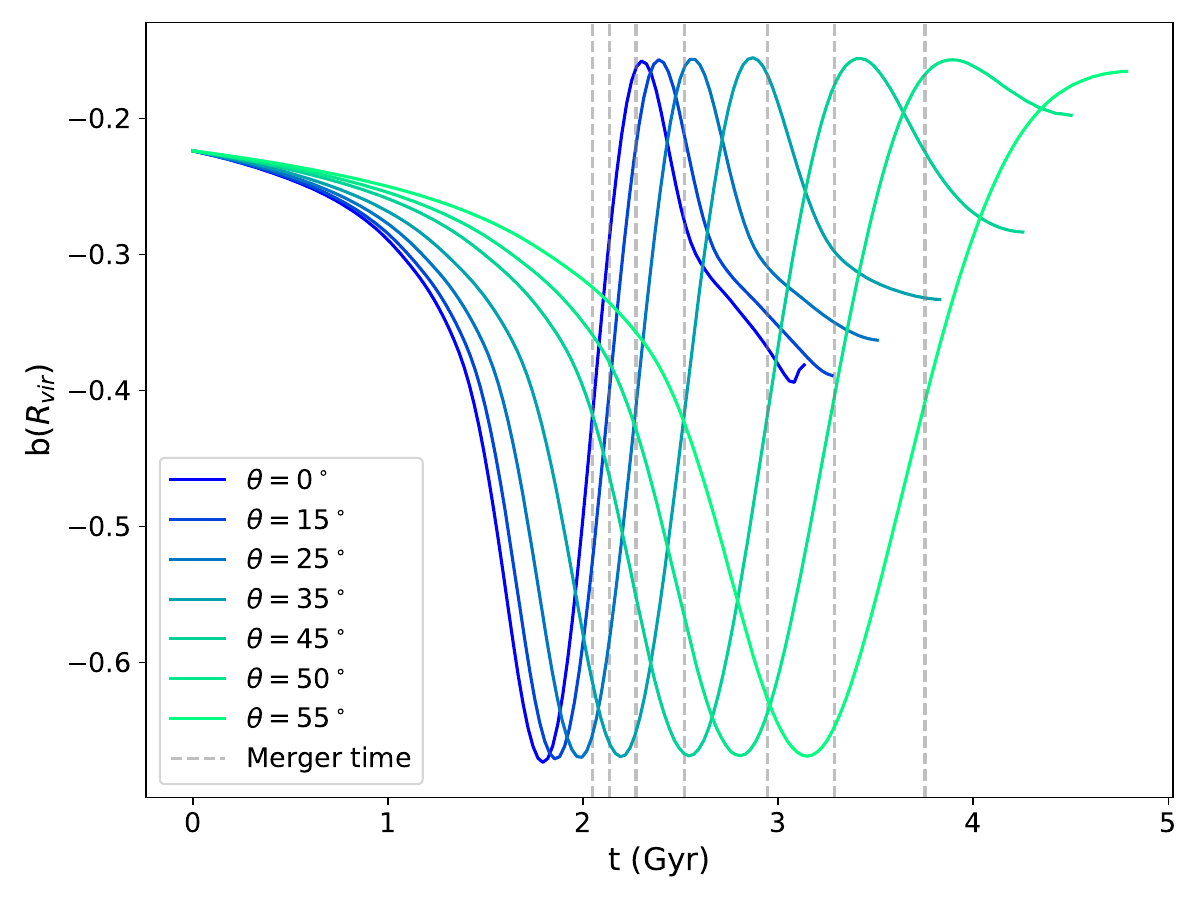}
  \caption{Outcome  of varying the orbital parameters on the merger mock simulations. \textit{Top panel}: Time evolution of the hydrostatic mass bias, evaluated at $R_{\mathrm{vir}}$, for different values of the mass ratio $q$. The impact parameter is fixed to $p=0$, and the  horizontal axis represents the time elapsed since the beginning of the test. The vertical dashed line represents the moment at which the infalling halo centre enters the virial volume of the host. \textit{Bottom panel}: Same evolution as the top figure but with a fixed mass ratio of $q=0.3$ and a varying impact parameter.}
  \label{mocktest} 
\end{figure}

In the panels of Fig.~\ref{mocktest} we show, respectively, the time evolution of the mass bias as a function of the mass ratio, $q$, for a fixed impact parameter ($p=0$, that is, a fully frontal collision), and as a function of the initial angle, $\theta$, between the initial velocity and the line that connects both clusters centres, which is related to the impact parameter as ${p=|\vec{r_\mathrm{ini}}||\sin\theta|}$, for a fixed mass ratio ($q=0.3$). 
As described by previous studies of single mergers \citep[e.g.,][]{Nelson_2012, Bennett_2022}, the temporal evolution of the mass bias shows a quite stable general shape: it becomes more negative towards the moment of the merger, it rises afterwards to reach a peak value and, then,  it decreases to a more stable value. 

As shown in the top panel of Fig.~\ref{mocktest}, this general trend  is preserved when varying the mass ratio of the merging halos. Namely, as the mass ratio increases --meaning the infalling halo is more massive--, the bias becomes systematically more negative and the central dip deepens, reflecting a stronger disruption of the HE condition. Furthermore, the timing of the main features, that is,  the location of the dip and the subsequent peak, vary for each test setup. This behaviour indicates that, as the mass ratio increases, and therefore, the relative velocity of the infalling cluster,   the dynamical impact in the final system is stronger and the and mass bias variations are advanced in the timeline of the event.

Complementarily, as shown in the bottom panel, the variation of the initial impact parameter modifies the orbital characteristics of the merger. Specifically, as the impact parameter increases, the merger becomes more tangential, delaying the pericentric passage.
This effect translates into a global rightward shift in the temporal evolution of the bias: both the main dip and the subsequent peak occur later and become broader in time. Despite this temporal displacement, the overall trend of the bias evolution is maintained and, thus,  it  preserves its characteristic dip-peak-stabilisation pattern.

These mock calculations, which disregard the compression of the ICMs, their heating due to adiabatic and/or shock compression, or any dynamical effect, serve only as a motivation and a baseline for the subsequent study. In the following sections, we aim to understand to what extent this evolution still holds in an actual physical setup, and how ICM thermodynamics modify it.

\subsection{Dynamical state of clusters and its connection with the hydrostatic mass bias} \label{relax state sec}

\begin{figure}
  \centering
  \includegraphics[width=\hsize]{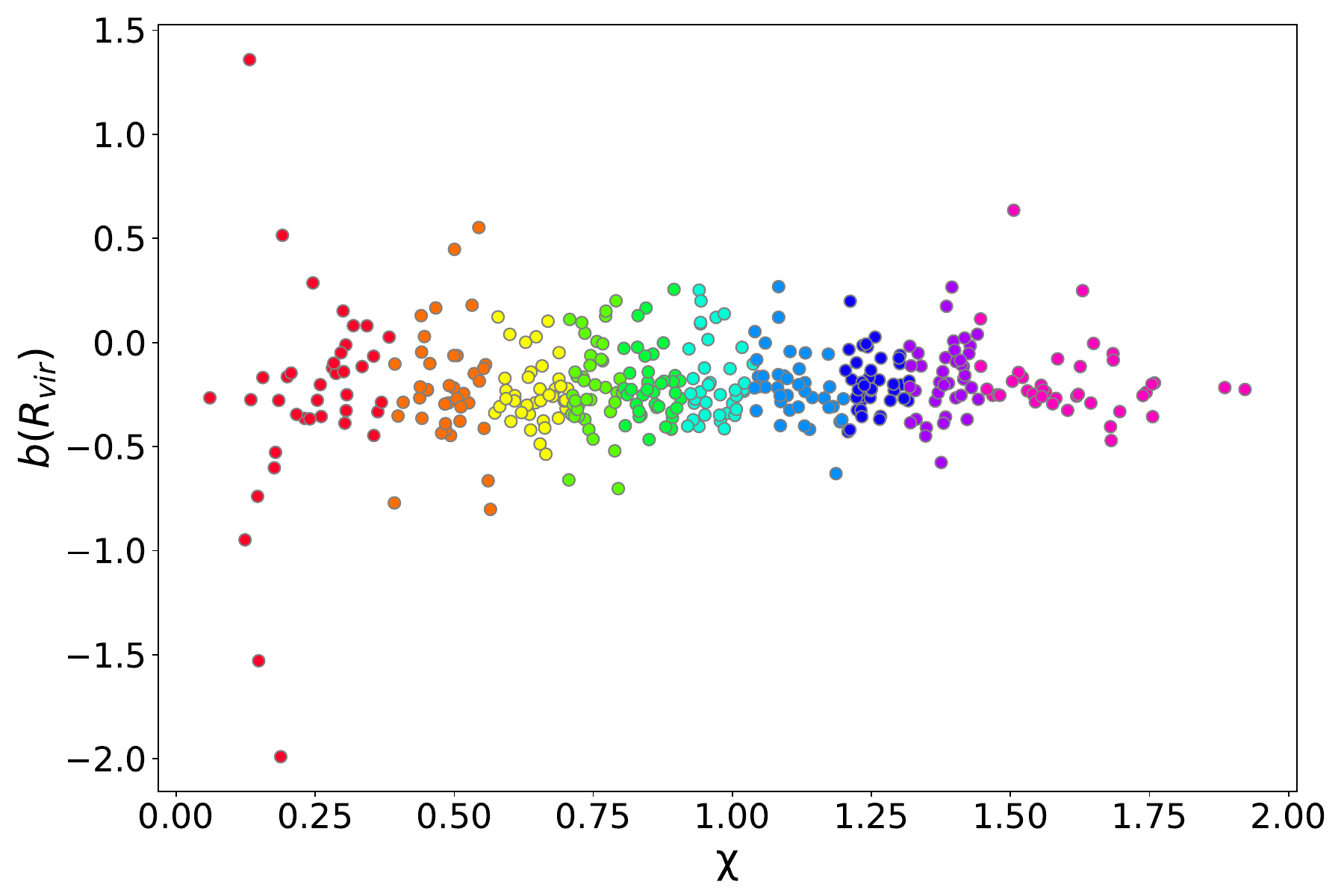}
  \caption{Hydrostatic mass bias evaluated at $R_\mathrm{vir}$ as a function of the combined dynamical state indicator, $\chi$, for all the clusters in our sample at $z=0$. Since the indicator has a continuous range of values, we set decile bins (each of a different color) with respect to the relaxation of galaxy clusters.}
  \label{relax bands z0} 
\end{figure}

Figure \ref{relax bands z0} shows the hydrostatic mass bias evaluated at the virial radius, $b(R_\mathrm{vir})$, as a function of the combined dynamical state indicator $\chi$, described in Sect. \ref{dyn state}, for all the clusters in our sample at $z=0$.

It is worth noting that the HE mass, and therefore the bias, depends exclusively on the local properties of the
density and temperature profiles, so the results we obtain at this specific radial location are only representative of the
behaviour of the mass bias at the virial radius. These results indicate that there exists a considerable scatter in $b(R_\mathrm{vir})$ within the different relaxation regimes.  However, an underlying trend is present, so that less relaxed systems (lower $\chi$) display more extreme values of the bias, either over- or underestimating the total gravitational mass by up to a factor of $2$. This behaviour demonstrates the fact that highly perturbed systems may be difficult to fit in the HE model with confidence.

Even though we are studying all galaxy clusters at the same cosmic epoch ($z=0$), this does not imply that they are all in the same evolutionary stage. As previously discussed, besides the significant impact of the cosmic environment,  galaxy clusters experience ongoing accretion and dynamical interactions, such as mergers and infall of smaller substructures, throughout their lifetimes.

Indeed, recent studies, both using high-resolution cosmological simulations \citep[e.g.,][]{Gouin_2021_connectivity} and using observational data \citep[e.g.,][]{euclid2025_connectivity}, have shown that the anisotropic structure of the environment surrounding galaxy clusters, quantified by their connectivity to cosmic filaments, plays a fundamental role in shaping their dynamical state and evolutionary history. Highly connected clusters tend to be more elliptical, accrete mass more rapidly, and are more likely to be dynamically unrelaxed. In contrast, weakly connected objects are typically more spherical, relaxed, and evolve more slowly. These two populations, early-forming relaxed clusters with low accretion rates and lately-formed, fast-accreting, unrelaxed clusters, exhibit distinct degrees of anisotropy in their density fields, directly influencing the stability and reliability of the HE assumptions. 

\begin{figure}
  \centering
  \includegraphics[width=\hsize]{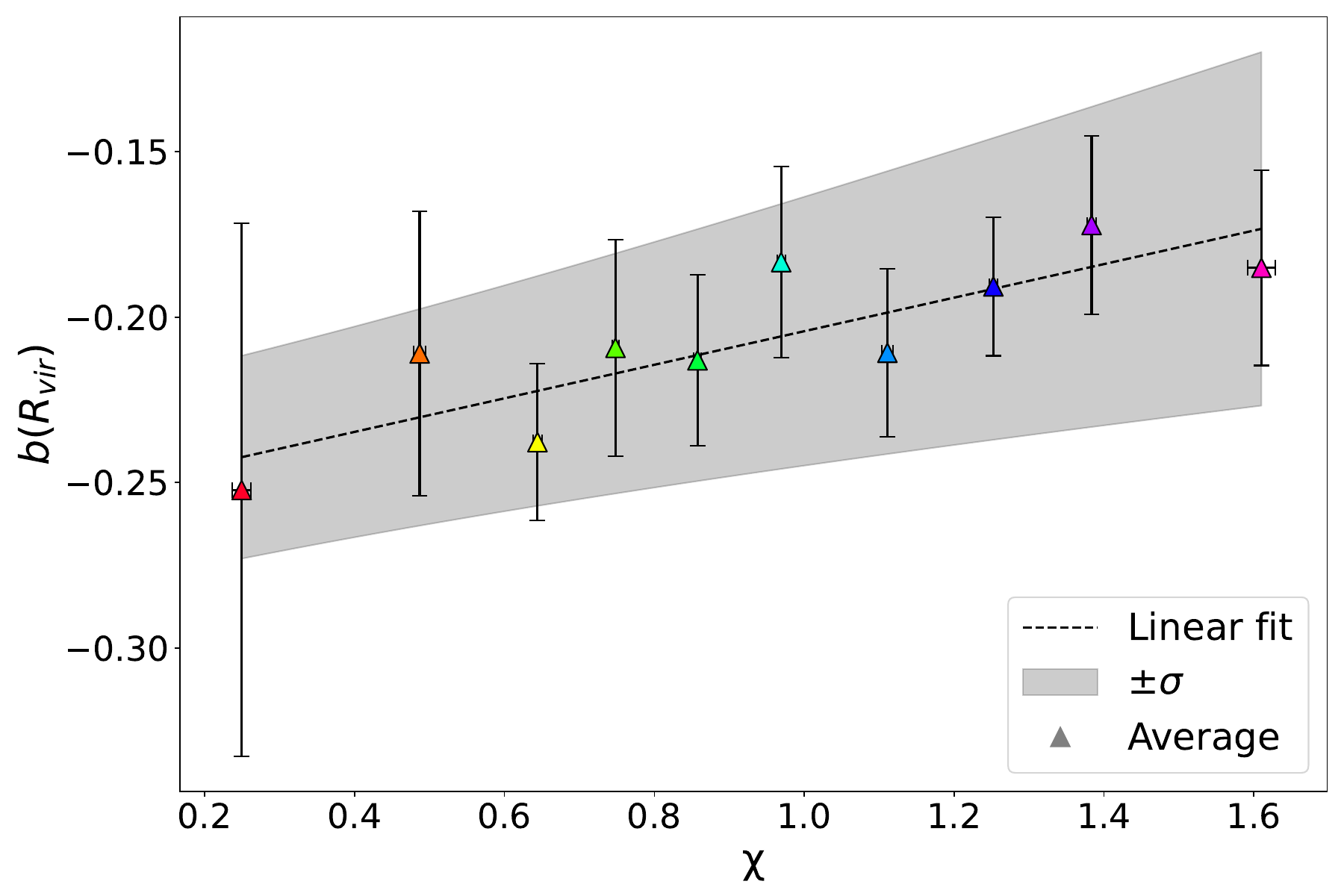}
  \caption{Average hydrostatic mass bias per dynamical state ($\chi$) bin, corresponding to the subsamples depicted in Fig~\ref{relax bands z0}. The error bars show the standard deviation within each bin, while the shadowed region corresponds to $1\sigma$ error on the linear fit over the average mass bias values (dashed line). }
  \label{avg relax bands z0} 
\end{figure}

In order to classify the clusters according to their degree of relaxation, we have set a decile binning for the values of the $\chi$ parameter. 
Despite the former mentioned intrinsic diversity, when considering the average mass bias per relaxation bin, as shown in Fig. \ref{avg relax bands z0}, a mild but consistent trend emerges: the more dynamically relaxed the cluster, the smaller the average mass bias at the given radius. This suggests a gradual tendency towards HE in relaxed systems. Specifically,  the linear fit, gives a parametrization of the bias such that ${b_{R_{\mathrm{vir}}}(\chi)=0.05^{(\pm 0.03)} \chi- 0.25^{(\pm 0.03)} }$, with a  p-value of $0.004$ and a  Spearman correlation of $0.81$.

Nevertheless, as shown in Fig.~\ref{relax bands z0}, a substantial dispersion in the mass bias values persists, not only towards underestimation (negative) but also towards positive values, implying cases where the HE-based mass measurement overestimates the true mass. This overestimation means that the inferred thermal pressure support surpasses the actual gravitational pull, possibly indicating the presence of transient density or pressure fluctuations, particularly in systems with a large infalling halo. This fact highlights how chaotic density and temperature profiles can become in extremely disturbed clusters. In some of these cases, the analysed evolutionary phase coincides with a particularly perturbed dynamical state, where the definition of the cluster centre may become ambiguous, leading to non-representative density and temperature profiles. Nevertheless, this behaviour is restricted to a few outliers and does not affect the overall trend.

Although at $z=0$ we do see a slight difference in the mass bias (in ensemble terms) between different levels of relaxation, on average the HE method underestimates the true mass, even in the most relaxed clusters, by about 20\% at the virial radius, a result consistent with previous findings \citep[e.g.,][]{Biffi_2016}. 

This picture renders the possibility to obtain cluster-by-cluster accurate determinations of $b$ based on dynamical state charactersations difficult; even though the relations in Fig. \ref{avg relax bands z0} could be useful first-order corrections if future work can study them in relation to simple ICM morphological parameters.

\subsection{The hydrostatic mass bias along galaxy clusters lifetime}
\label{mass bias evolution}

\begin{figure*}
  \centering
  \includegraphics[width=\hsize]{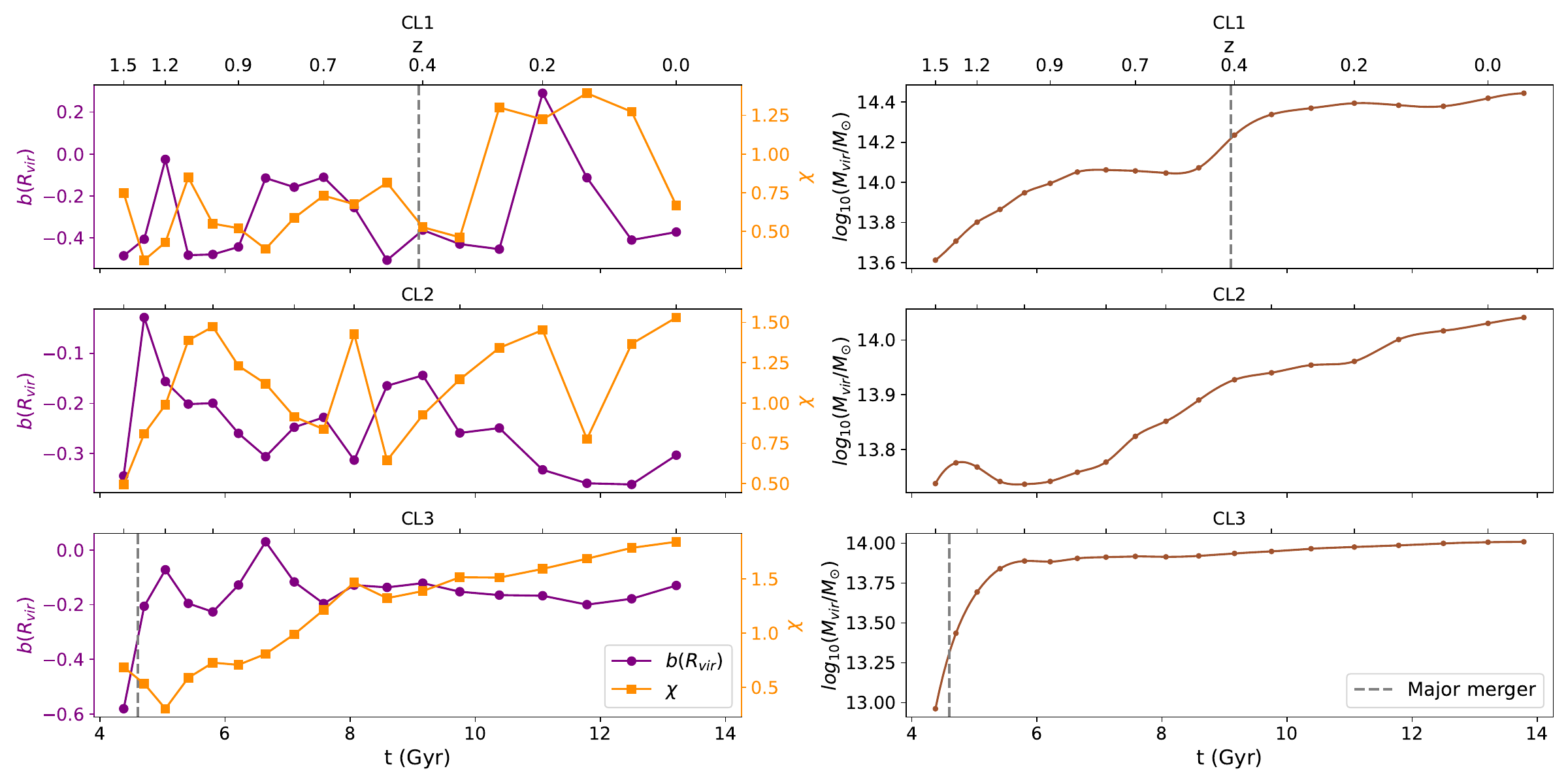}
  \caption{Prototypical  examples of hydrostatic mass bias evolution in relation to cluster assembly, for three particular clusters from our sample. {\it Left panels:} Hydrostatic mass bias evaluated at the virial radius, $b(R_\mathrm{vir})$ (purple), and of the dynamical state indicator, $\chi$ (orange). {\it Right panels:} Logarithmic mass growth history for the same period of time for the three clusters shown in the left panel. The vertical dashed lines indicate the time at which a major merger took place, in case there was any.}
  \label{bias X evolution 3 clusters} 
\end{figure*}

To illustrate the diverse evolutionary paths of galaxy clusters and their impact on their hydrostatic mass bias, Fig.~\ref{bias X evolution 3 clusters} shows the time evolution of the mass bias and the relaxedness parameter $\chi$ (left column), and  the virial mass (right column) for three representative galaxy clusters in our sample. In particular,  we have selected three properly resolved clusters with masses above $10^{14} M_{\odot}$ at $z=0$ and different mass growth histories. Specifically, we have chosen the cluster \texttt{CL1}, which has a major merger and  considerable mass accretion, the cluster \texttt{CL2}, with no major merger events but a substantial mass growth due to accretion, and lastly the cluster \texttt{CL3} (the one used to build the set-up in Sect. \ref{mock test}), which exhibits a strong merging event and a subsequent flattened mass growth curve.

In the top row of Fig.~\ref{bias X evolution 3 clusters}, we observe the cluster undergoing a relatively quiescent early evolution followed by a significant mass accretion event. The mass growth curve shows a plateau between 7 and 9 Gyr, after which the virial mass sharply increases due to a major merger event with a $1:2$ mass ratio. Tentatively, we can say that the mass bias and dynamical state approximately evolve in a coupled yet time-delayed fashion. 

In this situation, $b(R_\mathrm{vir})$ peaks around $\sim 1 \, \mathrm{Gyr}$ before $\chi$ does, as the value of the bias depends exclusively on the local state at $r \approx R_\mathrm{vir}$, while the dynamical state indicator $\chi$ accounts for the whole virial volume. As it is the case with the relaxation parameter, the relaxation of $b(R_\mathrm{vir})$ back to the baseline value is a long-lasting process, taking $ \tau_\mathrm{vir} \sim (2-3) \, \mathrm{Gyr}$ timescales. 
This behavior supports previous findings from \cite{Nelson_2012}, where post-merger relaxation is shown to be a multi-Gyr process, with thermalization and equilibrium restoration lagging behind the mass assembly. Interestingly, we can observe that the major merger, unlike the minor mass accretion events preceding it, triggers  a substantial increase in the mass bias after 2 Gyrs, even changing the sign (which means that the hydrostatic mass overestimates the real gravitational mass) for a short period of time and then decaying back to negative values.

The cluster depicted in the middle row of Fig.~\ref{bias X evolution 3 clusters} showcases a more chaotic evolutionary path. Here, both the bias and the relaxedness indicator fluctuate significantly without a clear correlation, albeit the amplitude of the bias variation is much more contained. This erratic pattern may reflect the influence of multiple smaller-scale accretion events or ongoing internal substructure interactions that continuously perturb the ICM (even mass losses as can be seen at around 5 Gyrs) rather than having a large merging event that temporarily dominates and shapes the behaviour of the cluster. In this case, the absence of a clear relaxation-bias relation aligns with the results from  \cite{Biffi_2016}, who reported that significant scatter exists in the hydrostatic mass bias-dynamical state relation, particularly for disturbed or dynamically young systems. This scenario reinforces the idea that not all clusters follow a neat evolutionary sequence, and stochastic processes may dominate over coherent relaxation trends.

Finally, the bottom row of Fig.~\ref{bias X evolution 3 clusters} shows the evolution of a cluster with a very different growth history. After an early, sharp major merger episode with a mass ratio of $2.5:1$ at around ${t = 4.5 \, \mathrm{Gyr}}$, the mass accretion rate flattens considerably, indicating a transition to a quiescent evolution phase. The  merger induces a pronounced spike in the mass bias, after which the system settles into a state of mild underestimation of the total mass ($b\approx -0.2$), which remains roughly constant despite the remarkable degree of dynamical relaxation. This suggests that non-thermal pressure components, such as bulk motions or turbulence, persist even in seemingly relaxed systems and sustain a residual bias \citep{Lau_2009, Nelson_2014}. Moreover, the lack of convergence to zero bias in a relaxed state challenges the simplistic view of HE validity and underscores the persistent role of microphysics and non-equilibrium processes in the ICM.

Overall, these three case studies underscore the fact that the time evolution of  the cluster mass bias is not solely determined by the mass growth, but also by the mode of accretion and the dynamical history. In the presence of major mergers, sharp changes in the mass bias and in the relaxedness indicator can be observed yet without a strong correlation (see Sect.~\ref{bias trend}), while in the absence of disruptive events, the system can stabilize into a dynamically relaxed configuration, where however the HE assumptions is still violated at the $\sim 20\%$ level.

\subsection{Evolution of the mass bias through a major merger event} 
\label{bias in merger}

As suggested by the results shown in previous sections, the comparison of galaxy clusters at a single snapshot in time may introduce substantial confusion, arising from the fact that individual clusters can follow distinct evolutionary trajectories, which are not adequately captured when temporal evolution is not considered.

Therefore, to ensure a consistent comparison across different evolutionary scenarios, we propose to focus our analysis on the most significant events in the evolution of galaxy clusters, that is, cluster major mergers.

To this end, the selection of merger events has been carried out following the methodology described in Sect.~\ref{catalogues and mtrees}, where the merger time is initially approximated as the moment when the secondary halo is inferred to cross the virial radius of the primary halo. However, due to the discrete nature of simulation outputs and the limited temporal resolution of this estimate, which only provides a temporal bracket for the merger time, we adopt a standardized definition of the merger time based on the local maxima of the mass accretion rate, $\Gamma(a)$, defined as:
\begin{equation}
    \Gamma(a)=\frac{\mathrm{d} \log M_{\text{vir}}}{\mathrm{d} \log a},
\end{equation}\label{gamma eq}
\noindent where $a=1/(1+z)$ is the cosmological expansion factor.

This definition of the mass accretion rate captures more accurately the physical growth of galaxy clusters and is only dependent on the determination of the main branch of the merger tree. 

Furthermore, we refine the final merger sample by excluding events that, despite potentially exhibiting high accretion values, do not present a local maximum in the $\Gamma(a)$ curve. Such cases,  characterized by a monotonically decreasing behaviour of the accretion rate, typically correspond to mergers obscured by extended periods of smooth mass accretion or by the superposition of numerous minor mergers.

From here on, in order to standardise the time scale at which the merging events occur and enable a proper comparison, we will use the  dimensionless time coordinate  $(t - t_{\text{merg}})/\tau_{\text{vir}}$, where the merger time $t_{\text{merg}}$ indicates the local maximum of $\Gamma (a)$ around the virial radius crossing time, thus marking  the origin of the temporal coordinate,  and the virial time $\tau_{\text{vir}}$ indicates the virial timescale at $t=t_\mathrm{merg}$, previously defined in Sec.~\ref{catalogues and mtrees}. 

\subsubsection{General trend during major mergers}\label{bias trend}

\begin{figure*}[ht]
    \centering
    
    \begin{subfigure}[b]{0.48\textwidth}
        \includegraphics[width=\textwidth]{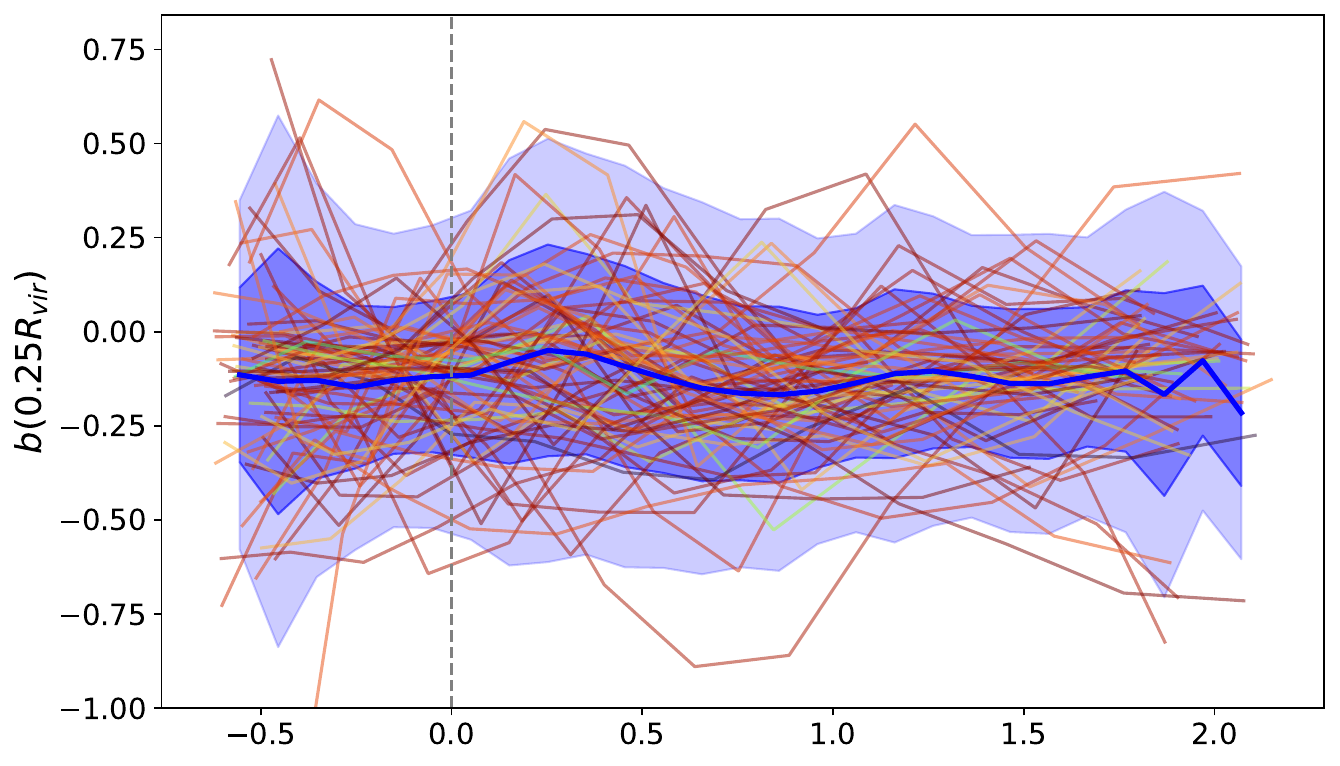}
    \end{subfigure}
    \hfill
    \begin{subfigure}[b]{0.48\textwidth}
        \includegraphics[width=\textwidth]{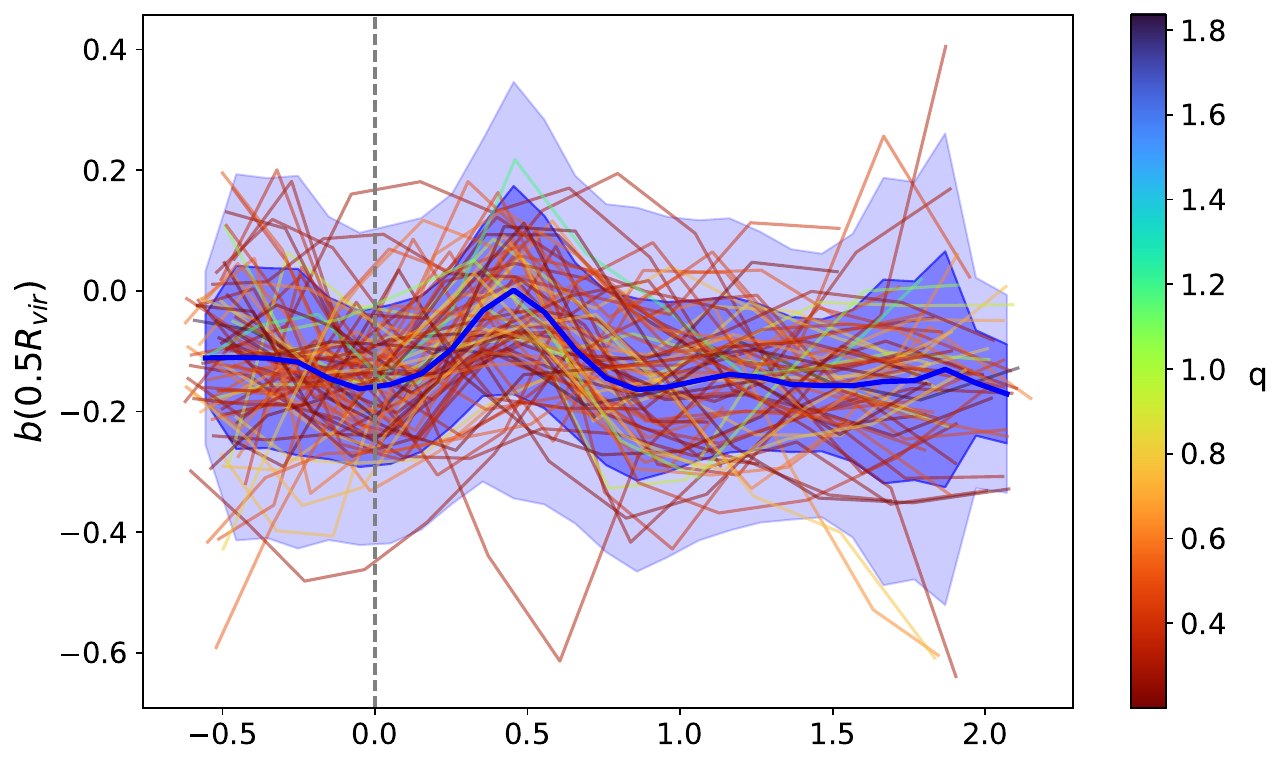}
    \end{subfigure}
    
    \vspace{0.4cm}
    
    \begin{subfigure}[b]{0.48\textwidth}
        \includegraphics[width=\textwidth]{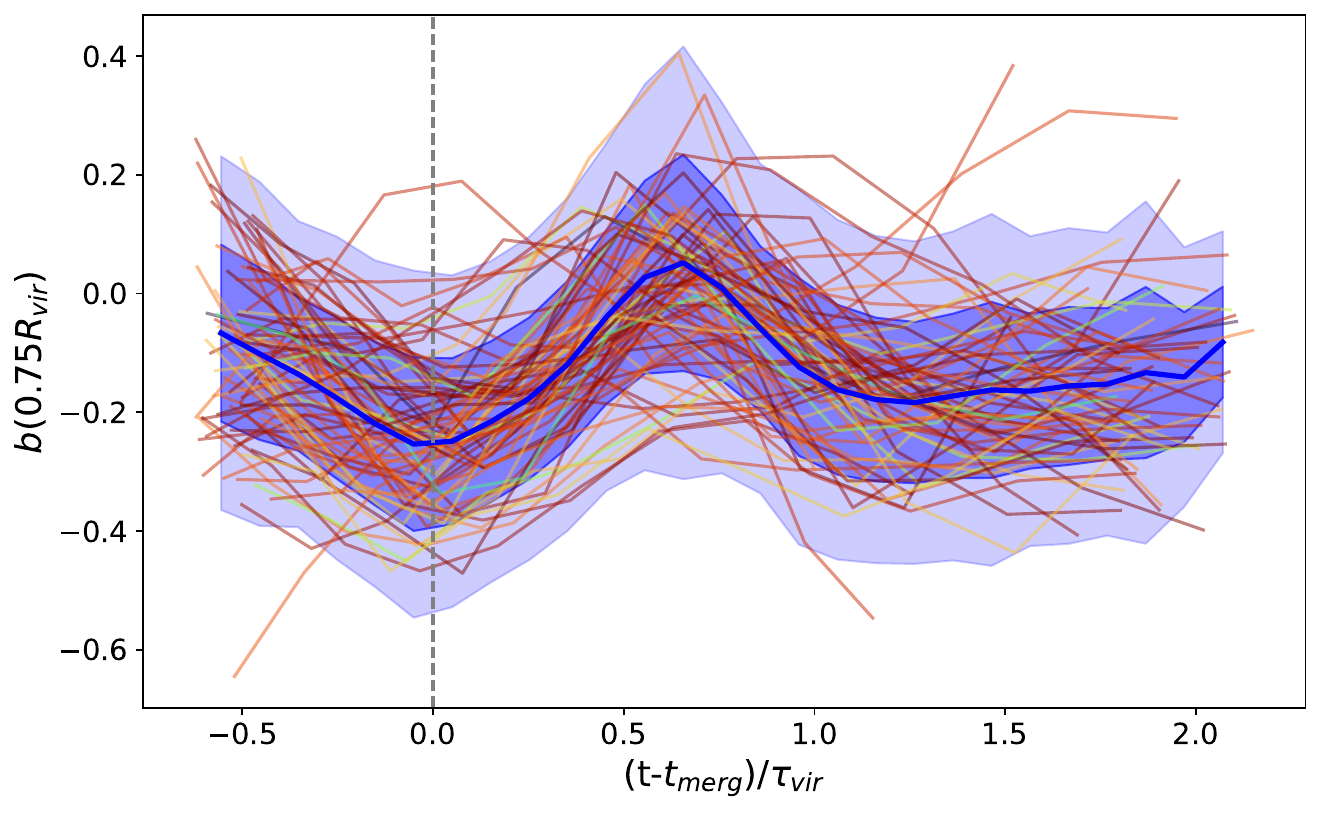}
    \end{subfigure}
    \hfill
    \begin{subfigure}[b]{0.48\textwidth}
        \includegraphics[width=\textwidth]{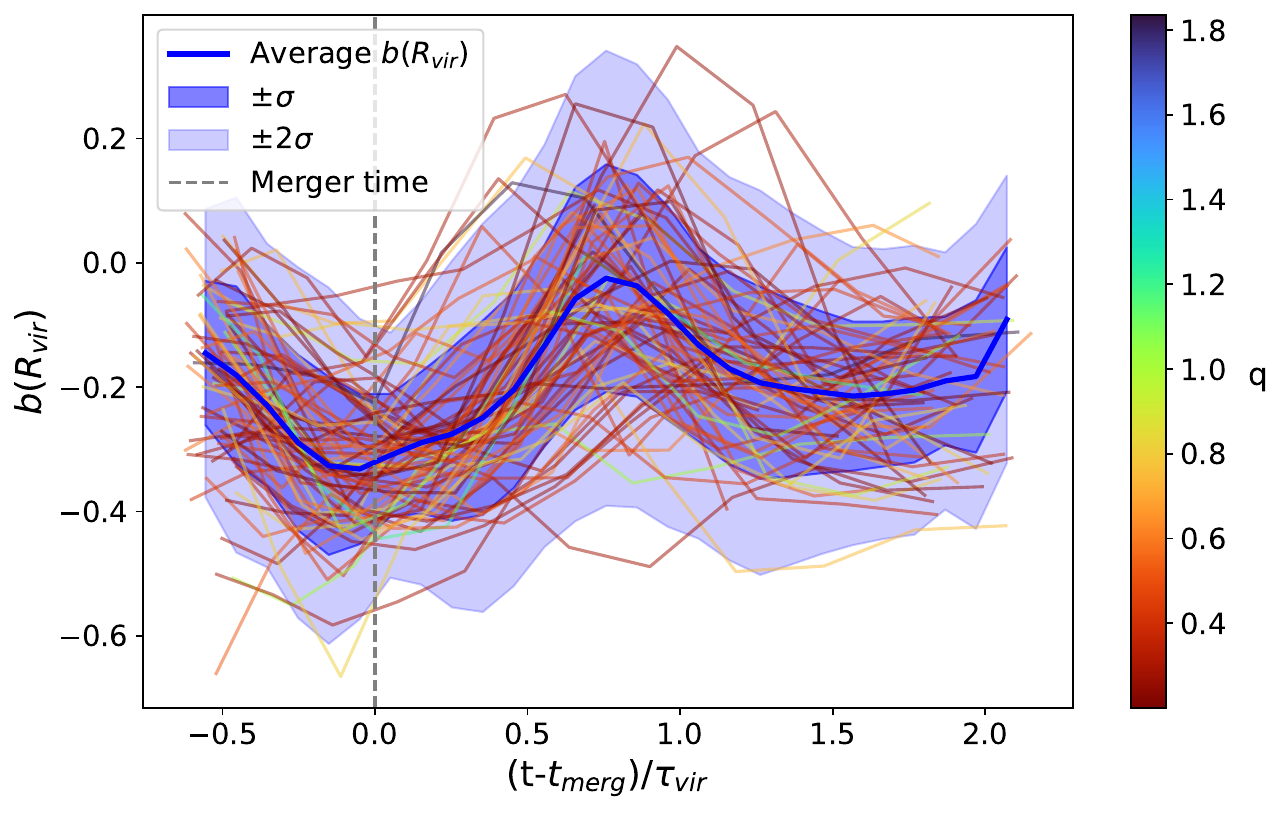}
    \end{subfigure}
    
    \caption{Standardised evolution of the hydrostatic mass bias, evaluated at several radii: $0.25R_{\mathrm{vir}}$, $0.5R_{\mathrm{vir}}$, $0.75R_{\mathrm{vir}}$ and $R_{\mathrm{vir}}$}, for the sample of isolated major merger events. The time coordinate is normalised to the virial timescale at the corresponding epoch as $(t - t_{\text{merg}})/\tau_{\text{vir}}$ and is vertically aligned with respect to the panels to facilitate comparison. The colourbar illustrates the magnitude of the merger mass ratio, $q$, and the average mass bias trend is depicted in blue with the corresponding $1\sigma$ and $2\sigma$ shadowed areas.
  \label{merger bias trend} 
    
\end{figure*}

Figure \ref{merger bias trend} presents the time evolution of the mass bias as a function of the normalized time, for the final sample of selected events. The study is conducted by evaluating the value of the mass bias at different fractions of $R_{vir}$ in order to observe how the HE mass bias evolves at different radii. The decision to use fractions of the virial radius instead of the more widely used overdensity-based radii is to avoid making comparisons that are affected by pseudo-evolution: since overdensity radii (e.g., $R_{2500}$, $R_{500}$, $R_{200}$) are defined with respect to the critical density, which evolves with redshift in a different way to $R_\mathrm{vir}$, the same $\Delta$ does not probe the same physical region of the halo at different epochs. For reference, the commonly used radii are roughly comparable to fixed fractions of the virial radius: $R_{2500}\sim 0.2 R_\mathrm{vir}$, $R_{500}\sim 0.5 R_\mathrm{vir}$ and $R_{200}\sim0.8 R_\mathrm{vir}$.  Each curve represents a different merger event colored according to the mass ratio $q \equiv M_{\text{infall}}/M_{\text{host}}$. The blue line of each panel shows the average bias trend and the dark and light blue shaded bands illustrate, respectively, the $1\sigma$ and $2\sigma$ confidence region across the sample at the corresponding radius. 

According to this figure, we find that the average bias evolution presents a flattened profile at small radii, maintaining the fluctuation of the average value throughout the evolution close to $-0.15$. This fact can be explained by how chaotic the profiles are at the core of galaxy clusters, added to the fact that the resolution in these very internal regions is not always optimal. Conversely, at intermediate and external radii we find a consistent and characteristic pattern associated with isolated major mergers, in agreement with findings from previous case studies \citep[e.g.,][]{Nelson_2012, Bennett_2022}. Specifically, the infall of the secondary halo into the host virial volume induces a pronounced decrease in the mass bias, reaching an average minimum value of approximately $b_{R_{\mathrm{vir}}}\approx- (0.35-0.3)$  near the time of virial crossing ($t_{\text{merg}}$) and smaller values, $b\approx- 0.25, -0.15$, for $0.75R_\mathrm{vir}$ and $0.5R_\mathrm{vir}$ , respectively.

This phase corresponds to the very onset of strong violations of the HE condition, as the infalling structure which just penetrated into the ICM starts to trigger turbulent motions as well as, potentially, a bow/merger shock and significantly distorts the density profile.
Following this trend, we observe a subsequent rebound in the mass bias, which reaches its peak roughly between half and one dynamical time after the virial-radius crossing, depending on the aperture considered. This occurs in spite of the fact that, during this period, the volume-filling fraction and intensity of turbulent motions tends to be maximal \citep{Vazza_2011, Valles-Perez_2021}, hence explicitly violating the HE assumption. In several cases, the bias even becomes positive, indicating that the hydrostatic assumption temporarily overestimates the true gravitational mass. This effect could be due to transient overpressures induced by shock heating or compression waves propagating through the ICM \citep[e.g.,][]{Rasia2012}. Subsequently, around $t\sim \tau_\mathrm{dyn}$ for the intermediate radii and $t\sim 1.5 \tau_\mathrm{dyn}$ at the virial radius, the system appears to reach a relaxed state, with the mass bias returning to values comparable to those observed prior to the merger. At this stage, the average bias reflects a mass underestimation of approximately $20\%$, followed by a mild increasing trend at the outer radii. This behaviour is consistent with the expected contribution of non-thermal pressure support commonly reported in cosmological simulations \citep[e.g.,][]{Biffi_2016}.

We find that the overall shape of the bias evolution remains consistent across different scales, indicating a coherent dynamical response of the cluster. A slight hint of that modulation can even be discerned at the innermost radius. The characteristics of this modulation, however, vary 
systematically with radius. The timing of the bias peak shifts outward: it appears slightly before $0.5\,\tau_{\mathrm{vir}}$ at $0.5\,R_{\mathrm{vir}}$, just after $0.5\,\tau_{\mathrm{vir}}$ at $0.75\,R_{\mathrm{vir}}$, and around $0.75\,\tau_{\mathrm{vir}}$ at the virial radius. The modulation amplitude also increases with radius, remaining closer to zero in the inner region but reaching larger deviations from the baseline at $0.75\,R_{\mathrm{vir}}$ and $R_{\mathrm{vir}}$. In addition, the duration of the perturbation grows with radius: the inner regions relax earlier, whereas at the virial radius the bias remains disturbed for up to $0.5\,\tau_{\mathrm{vir}}$ longer than at $0.5\,R_{\mathrm{vir}}$. This radial dependence reflects the propagation of the merger perturbation through the ICM: the core and inner regions respond more promptly and moderately, recovering quickly, 
while the outer region experiences a delayed, stronger, and more persistent disturbance.

\begin{figure}
  \centering
  \includegraphics[width=\hsize]{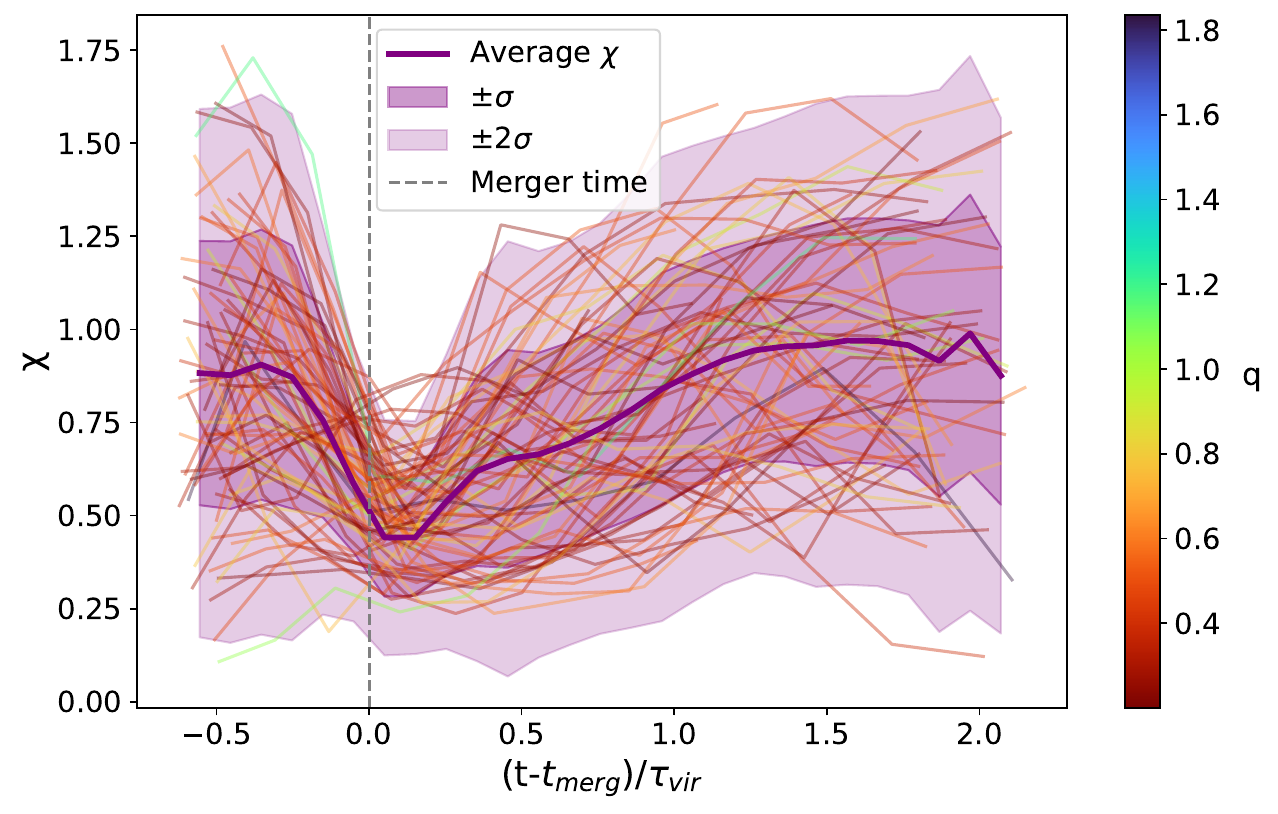}
  \caption{Evolution of the dynamical state indicator during the major merger event for the selected events. Colourbar shows the merger mass ratio, $q$,  the time coordinate is normalised to the virial time, $\tau_\mathrm{vir}$,  and the average $\chi$ trend is depicted in purple with the corresponding $1\sigma$ and $2\sigma$ shadowed areas.}
  \label{merger dyn indi curves} 
\end{figure}

For completeness, Fig.~\ref{merger dyn indi curves} presents the time evolution of the dynamical state indicator, $\chi$, for the same sample of mergers shown in Fig.~\ref{merger bias trend}. As expected, we observe a dip in the indicator near the merger epoch, reflecting the strong deviation from dynamical relaxation caused by the morphological and kinematic asymmetries introduced by the infalling cluster. In this situation, the centre of mass, the minimum of the potential, and the density peak become increasingly misaligned as the merging cores interact, which significantly reduces the relaxation indicator. 
The subsequent increase in the average trend in $\chi$ indicates the onset of the relaxation phase, during which the cluster progressively regains a more symmetric morphology and exhibits a reduction in bulk gas motions, consistent with trends reported in previous studies \citep[e.g.,][]{Cui_2018, Pearce_2019}.

As shown in Figs.~\ref{merger bias trend} and \ref{merger dyn indi curves}, while the average time evolution of the mass bias and the $\chi$ indicator is quite robust, a noticeable event-to-event spread persists in both quantities. This dispersion could be tentatively attributed to parameters such as the impact parameter or the mass ratio of the merger. As shown in the mock test from Sect.~\ref{mock test}, a larger impact parameter tends to delay the interaction and create more tangential orbits, stretching the timescale of the perturbation and  broadening the peak and dip in the bias curve. Similarly, mergers involving more massive infalling halos tend to produce larger deviations in the mass bias. However, our analysis reveals no strong correlation between the mass ratio $q$ and the peak amplitude of the bias, nor between the impact parameter and the timing of the event. This suggests that additional factors, such as the dynamical interaction between the secondary halo and the ICM of the host and the presence of pre-existing substructures, may significantly influence the evolution of these events, introducing further complexity into the physical processes governing cluster mergers.

In this context, it is worth noting that the temporal evolution of the dynamical state indicator behaves differently from that of the hydrostatic mass bias. The  composite indicator, as well as the involved individual indicators, generally exhibit a nearly symmetric evolution around the merger time, approximately satisfying $X(t_\mathrm{merg}-\Delta t) \approx X(t_\mathrm{merg}+\Delta t)$ for small $\Delta t$. This symmetry reflects the gradual disruption of dynamical equilibrium as the merger begins, followed by a comparable relaxation phase.

Conversely, the evolution of the hydrostatic mass bias at all radii is intrinsically asymmetric with respect to $t_\mathrm{merg}$. The bias deepens near the virial radius crossing time, corresponding to a stronger mass underestimation, and subsequently overshoots, reaching positive values before slowly returning to a quasi-stable level. As a consequence, a given bias value can occur at distinct dynamical stages, explaining the lack of correlation detected between the instantaneous degree of relaxation and the hydrostatic mass bias, glimpsed in Sect.~\ref{mass bias evolution}, and motivating our approach of explicitly following the time evolution of the hydrostatic mass bias throughout the merger process.

\subsubsection{A functional prescription for observational mass bias calibration} \label{merger trend fit}

Forthcoming X-ray facilities, such as NewAthena, are expected to deliver surveys detecting large samples of galaxy clusters with unprecedented depth \citep{Castellani_2024}. These will build upon current, state-of-the-art samples (X-COP, \citealp{Eckert_2017}; CHEX-MATE, \citealp{Chexmate_2021}). Meanwhile, in the optical, instruments like Euclid can provide valuable constraints on the dynamical state and merging history of clusters \citep{euclid2025_connectivity}. In this context, the calculation of cluster masses from observational data using HE arguments seems to be crucial, although the use of the traditional approach based on a fix bias correction appears to be rather poor.

In this regard, the results presented in the present work could dramatically improve the mass estimation based on HE hypothesis. Thus, 
we present  simple functional forms that model the evolution of the mass bias at different radii as a function of time in the vicinity of a major merger event based on the results described in previous subsections. This fit, calibrated on the average trend of the hydrostatic mass bias  across our sample, could allow an observational estimation of bias based on the relative time to the merger.

For this purpose, the fitting is restricted to the time interval encompassing a single oscillation period of the mass bias, centred around the merger time where the main modulation occurs. This choice is motivated not only by the domain covered in this work, but also by observational limitations. Outside this interval, the bias modulation due to the merger becomes less pronounced or increasingly degenerate with other dynamical disturbances, and observationally, determining the precise time since the merger is challenging due to the limited precision of morphological or substructure-based diagnostics. Consequently, this range represents a reasonable compromise between model applicability and observational feasibility.

We found that the observed evolution of the bias  can be fitted to a sinusoidal modulation with the following shape: 
 \begin{equation}
    b_r(\tau)=A+B~\sin(C~\tau + D),
    \label{sin fit}
\end{equation}
\noindent where the fitted parameters $A$, $B$, $C$ and $D$ correspond, respectively, to the baseline, amplitude, frequency and phase.

Particularly, for the virial radius we obtain: 

\begin{equation}
    b_{ R_{\mathbf{vir}}}(\tau)= -0.19^{(\pm 0.03)}  + 0.13^{(\pm 0.05)}  \sin\left(3.5^{(\pm 0.5)}   \tau {-1.5}^{(\pm 0.4)}\right)\, ,
\end{equation}

\noindent where $\tau\equiv (t - t_\text{merg})/\tau_\text{vir}$.

\begin{figure}
    \centering
  
  \includegraphics[width=\hsize]{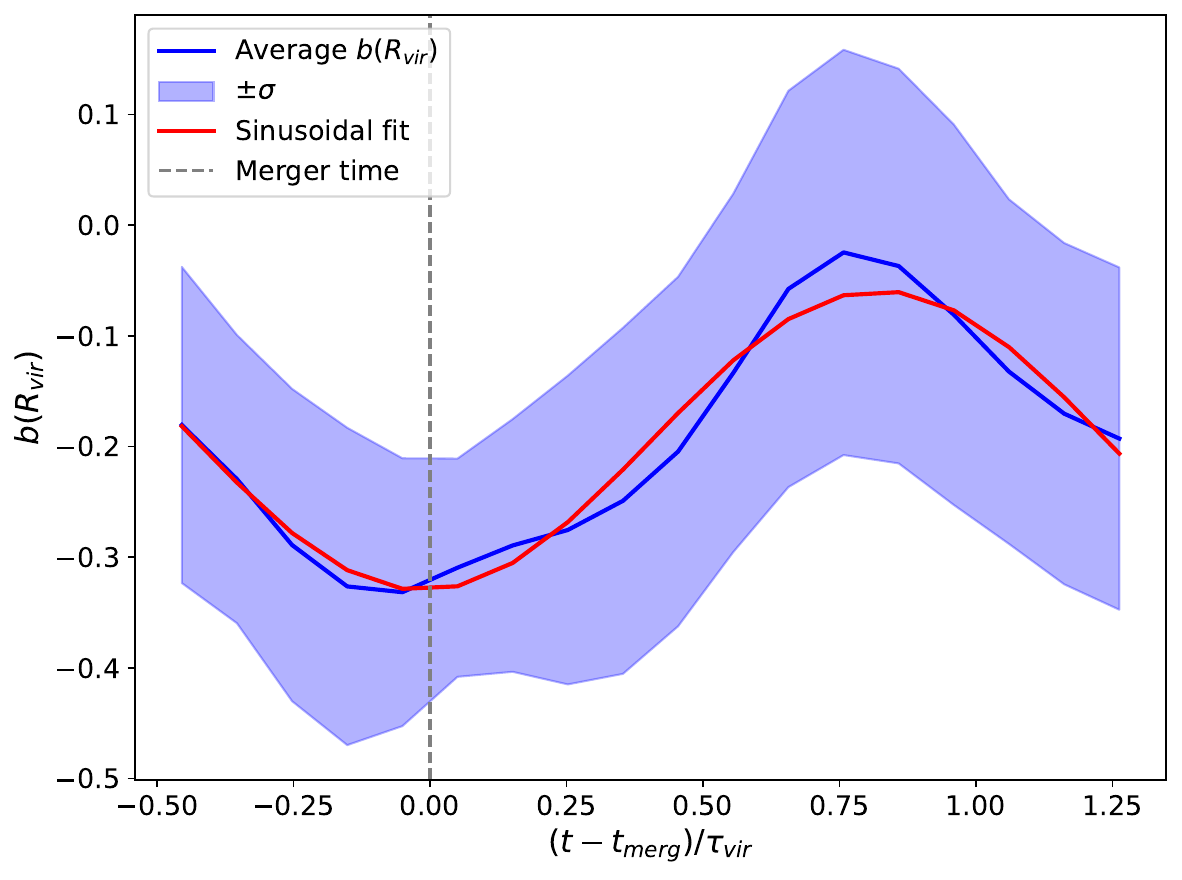} 
    \caption{Sinusoidal fit (red) over the average mass bias trend evolution evaluated at $R_{\mathrm{vir}}$ during major mergers (blue), with the corresponding $1\sigma$ shadowed area.}
    \label{sinus fit}
\end{figure}

\begin{figure}
      \centering
      \includegraphics[width=1\linewidth]{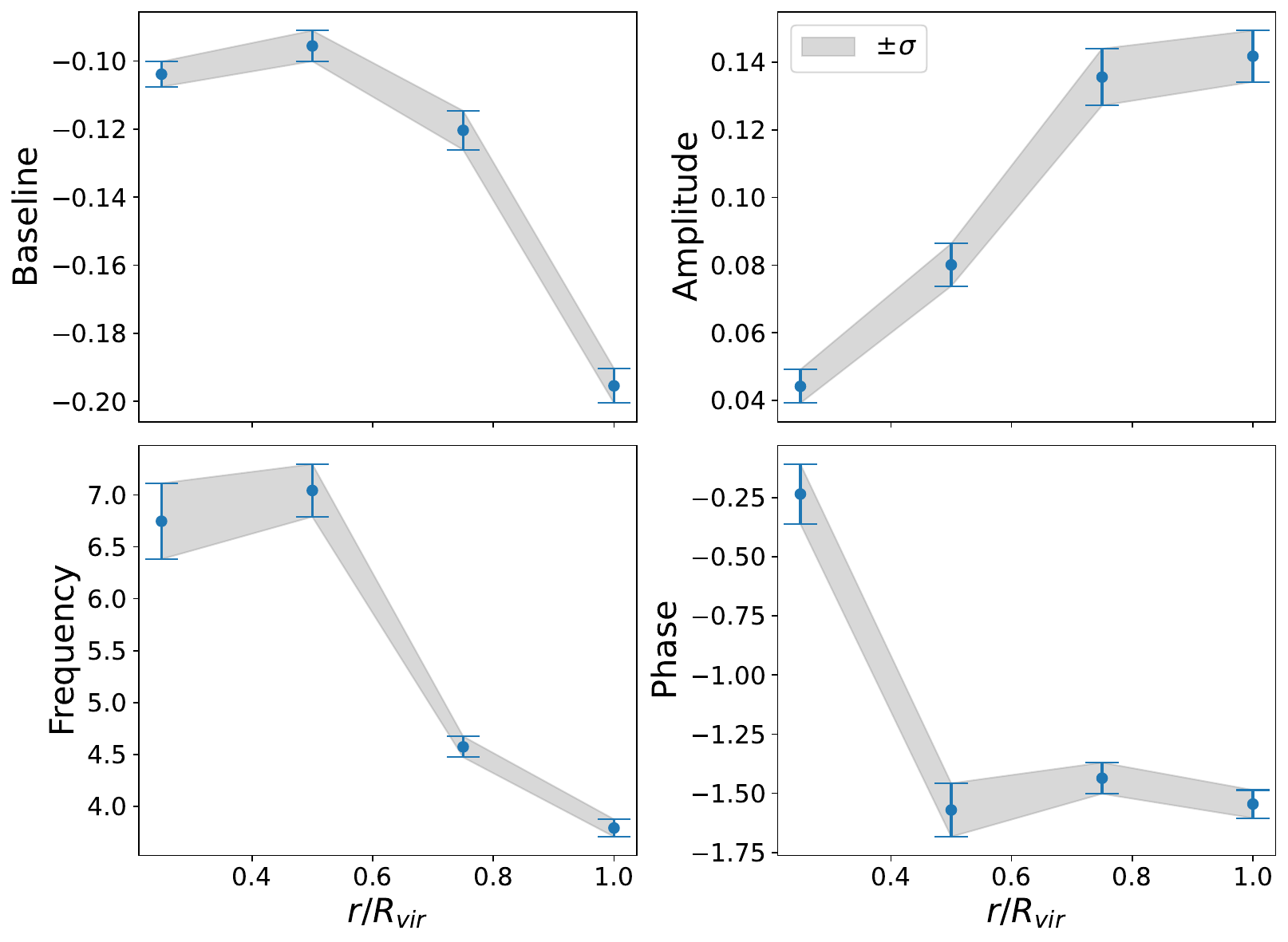}
      \caption{Radial variation of the  parameters of the sinusoidal fit (Eq. \ref{sin fit}) for the  mass bias trend within the main oscillation. Each scatter point corresponds to the respective fitted parameter value with its statistical error bars at a given radius.}
      \label{parametres_trend}
  \end{figure}

This fit can be applied to the evolution of the bias at different radii, with the time interval adjusted to match a single oscillation period. Across the four cases shown in Fig.~\ref{merger bias trend}, the radial dependence of the fitting parameters shown in Fig.~\ref{parametres_trend} supports the previously proposed idea: the hydrostatic mass bias oscillation exhibits lower amplitude and higher frequency at smaller radii. Consequently, in the inner regions of the clusters, the imprint of the merger on the density and temperature profiles is weaker and dissipates more rapidly. In addition, as we approach larger radii, the baseline shifts toward more negative values, leading to a more negative average bias overall. Although not reported in the figure, extending the fit to larger radii, we found that the amplitude trend stops and begins to decrease, so that the effects of the merger on the bias are diluted outside this boundary as the radius increases.

However, given the difficulty of accurately determining  $t_{\mathrm{merg}}$ observationally, we offer a more practical approximation to estimate the hydrostatic mass bias based on these results. Thus, we divide the merger event into three pseudo-constant stages: during the substructure infall, just after the pericentric passage, and  when the effects of the merger have dissipated (namely, pre- and post-merger). Please note that with this division, we encompass the pre-merger and post-merger periods, as they are qualitatively equivalent for the purposes of this study.  With this, we separate mass bias variation in three regions. This sequence corresponds first to the drop in mass bias caused by the entry of the subhalo into the host, followed by the subsequent increase driven by the merger shock and thermalisation, and finally the decline back toward the pre-merger level: 
$b_{0.5R_{\mathrm{vir}}} \approx \{-0.15,\, 0,\, -0.1\}$ 
and 
$b_{R_{\mathrm{vir}}} \approx \{-0.3,\, 0,\, -0.2\}$, 
corresponding respectively to the phases during the first infall, after the pericentric passage, and in the pre- and post-merger state.

Note that this prescription is just indicative and the characteristic time scales vary for different radii, as we have shown in Fig.~\ref{merger bias trend} and Fig.~\ref{parametres_trend}.

Despite its simplicity, this model provides a useful and flexible framework to incorporate the merger-phase-dependent modulation of the mass bias in observational analyses. By enabling a time-varying correction rather than a fixed value, it allows for more accurate hydrostatic mass estimates in dynamically active clusters. This refinement can help mitigate systematics in cluster-based cosmological studies, offering a practical improvement over the conventional, non-evolving bias correction schemes.\\

\subsubsection{Exploring the main drivers of the mass bias evolution}
\label{bias trend causes}

\begin{figure*}  
\centering

\begin{minipage}[b]{1.045\textwidth}
\hspace{-0.5cm}
\includegraphics[width=\linewidth]{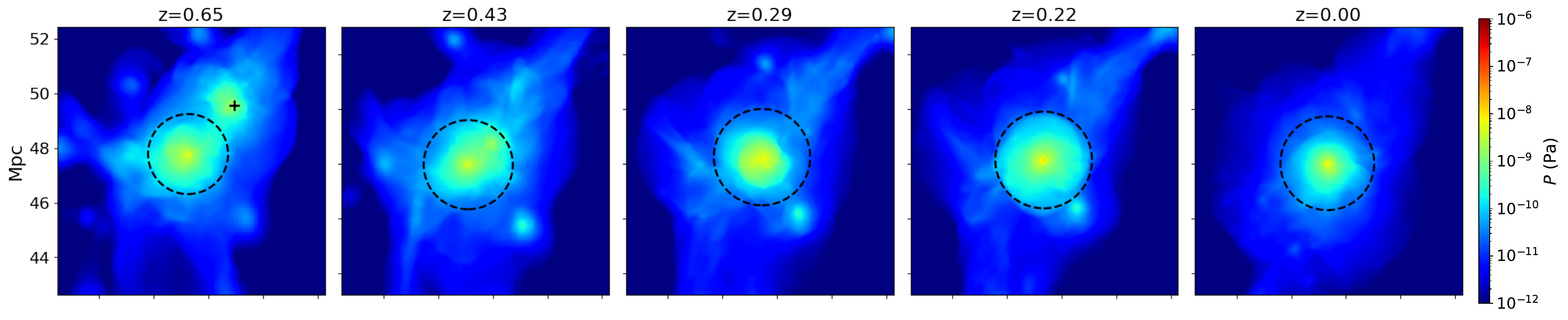}
\end{minipage}

\begin{minipage}[b]{1.045\textwidth}
\hspace{-0.5cm}
\includegraphics[width=\linewidth]{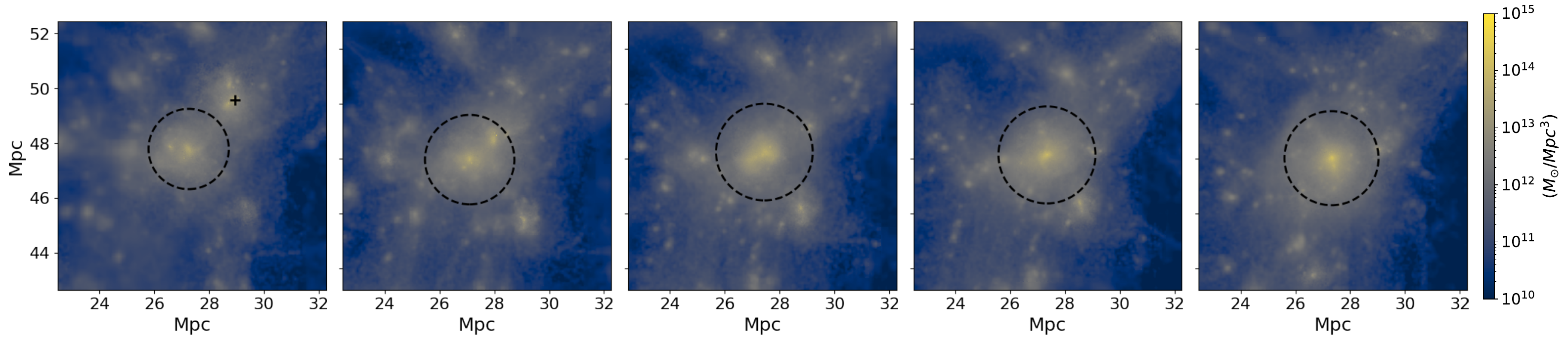}
\end{minipage}

\begin{minipage}[b]{0.97\textwidth}
\hspace{-0.5cm}
\includegraphics[width=\linewidth]{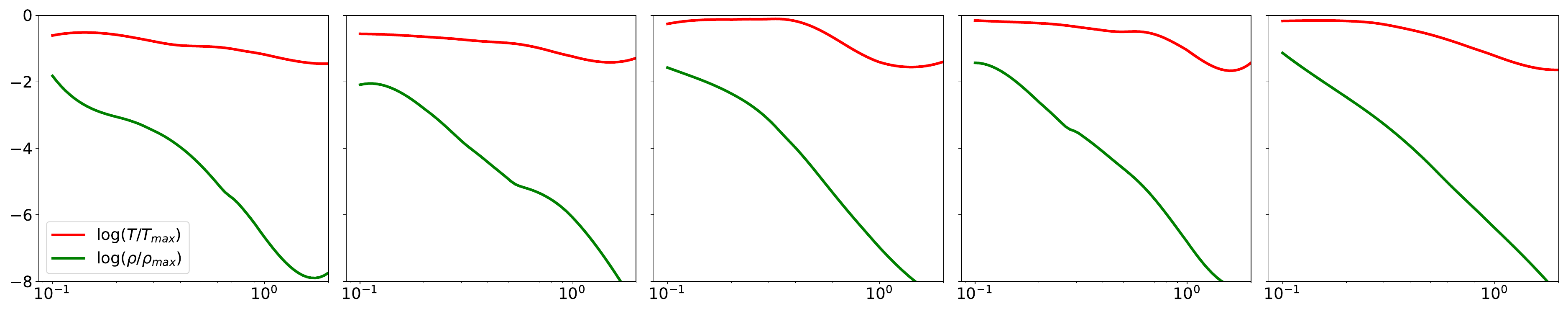}
\end{minipage}

\begin{minipage}[b]{0.98\textwidth}
\hspace{-0.55cm}
\includegraphics[width=\linewidth]{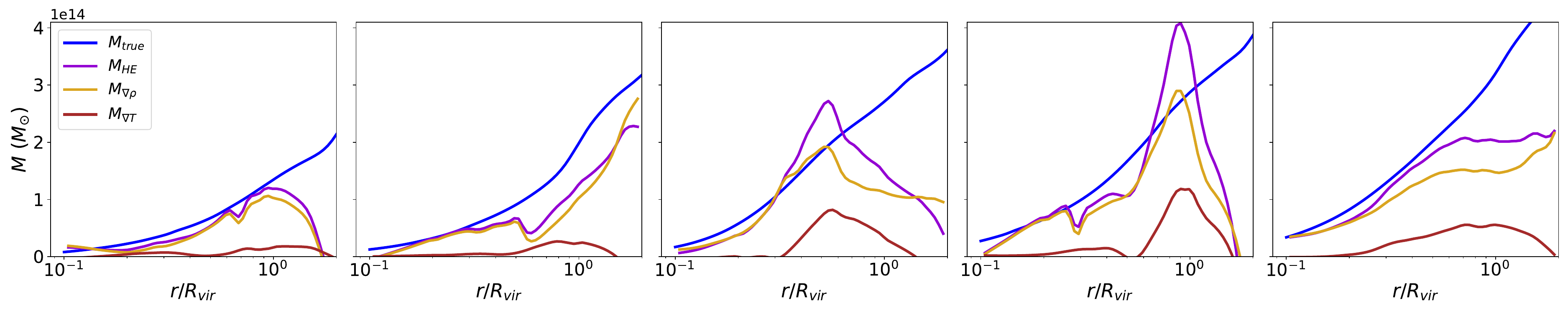}
\end{minipage}

\caption{Temporal evolution of a merging event  between two galaxy clusters with a mass ratio of 1:2. {\textit{Top panels}: Variation of the projected pressure map during the event. \textit{Second row panels}: Evolution of the projected total density. In these panels, within the 2D projected maps the virial radius of the host halo is marked with a dashed circle, whereas the center of the infalling halo, while being outside of the host, is given by a black cross. \textit{Third row panels}: Evolution of the logarithmic temperature (red) and density (green). The thermodynamic variables have been normalised to the maximum of their evolution to allow a visual comparison. \textit{Bottom panels}: Evolution of the true mass radial profile (blue), the hydrostatic mass profile estimate (purple) and, according to Eq.~\ref{mHE formula}, the individual contributions to the hydrostatic mass profile from density (yellow) and temperature gradients (brown).}}
\label{fig:multirow_merger}
\end{figure*}

In this section, we focus on a single major merger event to study in more detail the physical mechanisms responsible for the mean trend observed in Fig.~\ref{merger bias trend}. 

To this end, the system under study is cluster \texttt{CL1}, one of the most massive galaxy clusters at $z = 0$, with a pre-merger virial mass of ${M_{\text{vir}} = 2.78 \times 10^{14} \, M_{\odot}}$ and a virial radius of ${1.72 \, \mathrm{Mpc}}$. For this cluster, in order to capture both the immediate and longer-term effects of the merger, which has a mass ratio of 1$:$2, we analyse the period of time between $z = 0.65$ and $z = 0$. Notably, this cluster has not suffered any additional major merger within the analysed redshift interval, implying that it has remained free of such events for approximately 5 Gyr (see top panels of Fig.~\ref{bias X evolution 3 clusters}). Nevertheless, its dynamical state prior to the merger is classified as perturbed, highlighting the complex and active evolution of galaxy clusters even in the absence of major merger activity \citep[][]{Biffi_2016}.

We examine the sequence of this major merger in Fig.~\ref{fig:multirow_merger}, where we show the 2D projected maps of pressure and total density fields for five different redshifts. In these maps, the center of the infalling halo is marked with a cross while it is out of the host virial volume, which is represented as a dashed-line circle. To complement these maps, the evolutions of the normalised temperature and density radial profiles are shown in the third row, and the bottom row presents the true and hydrostatic mass cumulative radial profiles of the host, as well as the separate contributions of the density and temperature terms in Eq.~\ref{mHE formula} to the mass estimate. 

We see that, even though the event is considered to be isolated in time and to be a one to one major merger, the cluster is embedded in a filamentary node in which the continuous stream of matter makes it difficult to clearly define its boundaries. In this regard, as already shown in previous works \citep[e.g.,][]{Musso_2018,Gouin_2021_connectivity,Kuchner_2021}, the connectivity of  galaxy clusters highly affects their evolution. For a highly connected  galaxy cluster, as opposed to an isolated  one, the mass accretion rate is higher, causing the halo to become more elliptical in shape as it accretes matter faster and, thus, maintaining a dynamically perturbed state. Consistently, the 2D maps of total density and gas pressure reveal the simultaneous infall of multiple smaller structures onto the host halo throughout the entire sequence. Nonetheless, the dominant contribution to the variation in the mass estimation of the host is expected to arise from the most massive of the infalling substructures.

At redshift $z = 0.65$, the infalling cluster is observed to approach the host from a filamentary structure, with its outer gaseous atmosphere beginning to overlap with that of the host halo. This interaction increases the pressure in the region between the two clusters, with respect to the surrounding outskirts of the host, leading to a flattening of  the density profile and, to a lesser extent, of the temperature profile in the outskirts of the halos, inducing a more pronounced decline in the hydrostatic mass estimate. This is observed as the change in the slope of the logarithmic profiles and in the mass profile in Fig.~\ref{fig:multirow_merger} at $r \gtrsim 0.9 R_\mathrm{vir}$, and illustrates the onset of the dip in $b$ observed in the global trend (Fig. \ref{merger bias trend}).
At this particular moment, the host halo is already classified as perturbed and  its hydrostatic mass  is underestimated by $\sim10\%$ at the virial radius. 
Then, moving to $z=0.43$, the secondary halo is about to cross the virial radius of the main cluster. This represents the closest snapshot to $t_\mathrm{merg}$ accessible for analysis. At this stage, the host cluster exhibits a markedly elliptical morphology due to the entrance of the infalling halo, corresponding to the maximum dynamical unrelaxedness (cf. Fig. \ref{bias X evolution 3 clusters}). The hot gaseous atmosphere of the merging halo induces a pronounced flattening of the density and temperature profiles, not only in the outskirts, where gas clumping and pressure inhomogeneities make the medium significantly less uniform \citep[e.g.,][]{Planelles_2017} and more prone to inconsistencies in the mass estimates, but also from $r \gtrsim 0.5 R_\mathrm{vir}$.

This results in a substantial reduction in the hydrostatic mass estimate, producing a bias as large as $b_{R_\mathrm{vir}}\sim -0.4$. However, supporting the previously presented idea, the innermost region of the cluster remains minimally affected by the disruption caused by the infalling halo, showing a minor deviation between the true and the hydrostatic mass profiles since, as mentioned, the greatest contribution to the variation in hydrostatic mass is due to the disturbance of the density profile by the infall of the secondary structure, which has not yet fully reached the inner region. This situation corresponds to the dip in the mass bias trend observed in Fig.~\ref{merger bias trend} at $t=t_\mathrm{merg}$, which for the smallest radius is hardly noticeable.

\begin{figure}
    \centering
    \includegraphics[width=\linewidth]{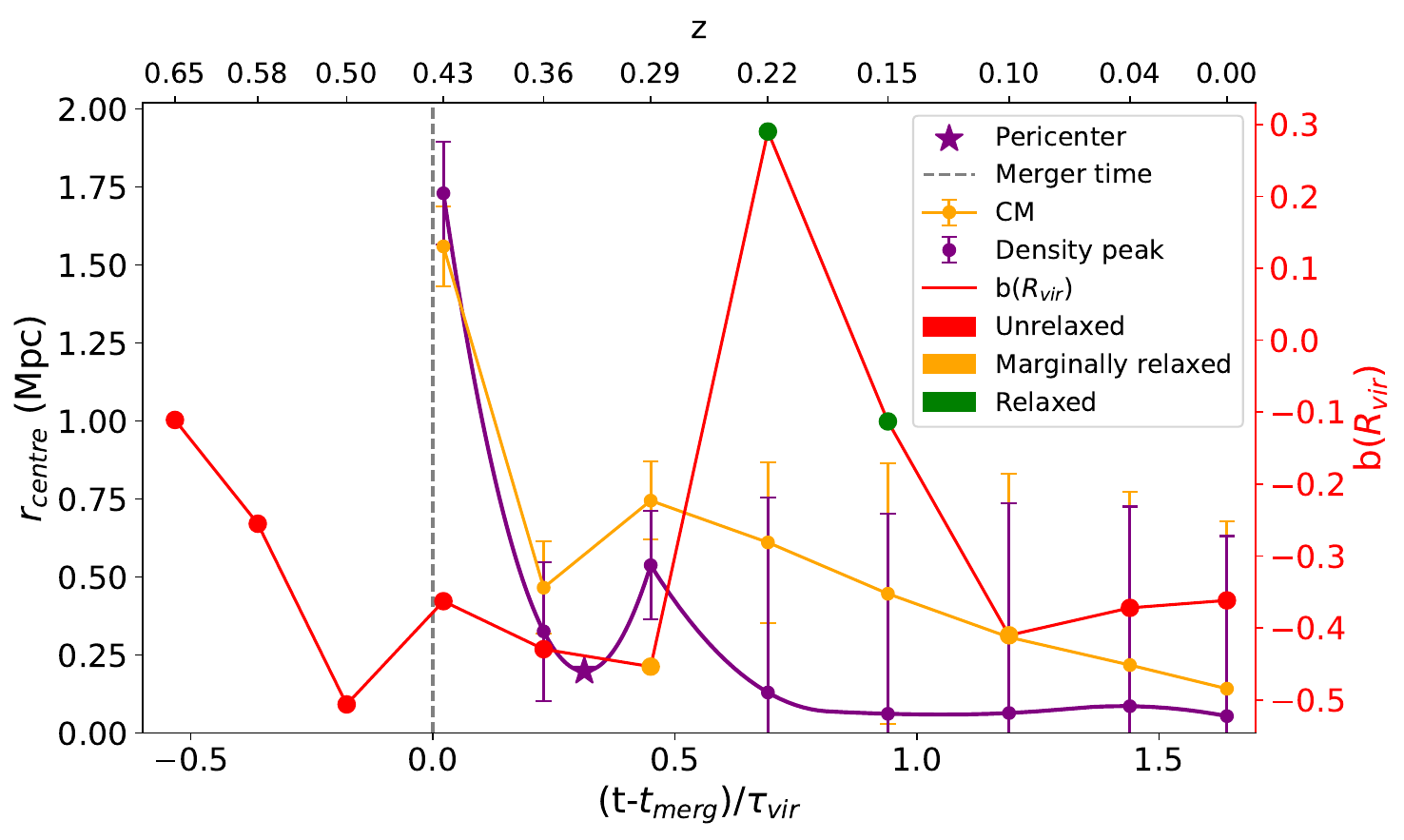}
    \caption{The red curve shows the hydrostatic mass bias (red line and right axis)  evolution during the major merger event for the galaxy cluster \texttt{CL1}. The color of the markers over the bias curve indicates the dynamical state according to the color code detailed in the legend. The connected dots with errorbars show the radial distance (orange line) between the center of mass (CM) and the density peak (purple line) of the merging halo with respect to the host (left axis). The star-shaped scatter point is the parabolic interpolation of the pericenter for the infalling orbit.}
    \label{fig:core_tracking}
\end{figure}

To complement this analysis, in Fig.~\ref{fig:core_tracking} we show the temporal evolution of $b_{R_\mathrm{vir}}$ together with the distance between the centres of the infalling and the host halo\footnote{To accurately track the infalling halo within the host virial volume, we followed the $1\%$ innermost pre-infall DM particles in phase space. To this end, we defined the phase-space distance as:
 
 $$d_{6D}(\mathbf{r}, \mathbf{v}) =  \left( \frac{|\mathbf{r} - \mathbf{r_{cm/peak}}|^2}{R_\mathrm{vir}^2} + \frac{|\mathbf{v} - \mathbf{v_{cm/peak}}|^2}{\sigma_\mathrm{v}^2} \right)^{1/2},$$

 \noindent where the subindex $\mathrm{cm/peak}$ stands for centre-of-mass or density peak quantities, respectively,  computed over all bound DM particles, and $\sigma_v$ is the three-dimensional velocity dispersion of the same set of particles.
 }.
 
 In addition, to show how the halo is being assimilated and diluted within the host, we consider the center of mass (CM) and the density peak of those innermost particles with their associated dispersion shown as error bars.

The panel at $z = 0.29$ in Fig.~\ref{fig:multirow_merger} corresponds to the first snapshot after the pericentric passage, indicating that the secondary halo has completed its initial infall. At this point, its velocity decreases as the gas of the infalling halo starts to be incorporated into the host's ICM, settling the secondary cluster into a less plunging, more stabilized orbit. As a result, the pressure map reveals the emergence of a quasi-spherically symmetric merger shock propagating outward from the centre of the main cluster. This shock front steepens the density  and temperature profiles as it propagates outwards, leading to a temporary  overestimation of the hydrostatic mass at each radii as shown in Fig.~\ref{merger bias trend}. However, the orbit of the secondary halo  is not strictly monotonically decaying, the subsequent passage after passing through the pericenter is relatively minor and does not significantly impact the bias curve. For comparison, a more complex multiple-merger scenario and its effects on the mass bias evolution are discussed in Appendix~\ref{appendix overlaping mergers}.

At $z = 0.22$, the merger-induced shock has propagated further outwards, generating a steep pressure gradient near the virial radius. This situation results in a maximum overestimation of the hydrostatic mass by approximately $30\%$ at the virial radius (peaking with $b\approx 0.35 $ at $\sim 0.9R_{\mathrm{vir}}$), occurring roughly $0.75\,\tau_{\text{dyn}}$ ($\sim$1.9 Gyr) after the merger event. At this stage, most of the material from the secondary halo has been disrupted and incorporated into the host, as evidenced by the substantial particle dispersion, reflected in the large error bars shown in Fig.~\ref{fig:core_tracking}.
Subsequently, after approximately 4.5 Gyr, the most prominent effects of the merger have vanished. During this phase, although the newly formed cluster begins to settle into a more relaxed configuration, the mass bias remains significantly negative ($b_{R_\mathrm{vir}}\sim -35\%$). This residual bias is expected to gradually become smaller as the gas in the outer regions thermalises and the bulk and turbulent motions diminish, thereby reducing the contribution of non-thermal pressure support.

This results are well in agreement with previous works by \cite{Nelson_2012} and \cite{Bennett_2022}. However, a noteworthy finding is that the temperature variation alone does not fully account for the magnitude of the observed bias deviation. As the radial profiles reveal, and as suggested by our toy model for a merger in Sect.~\ref{mock test}, the dominant contribution to the hydrostatic mass underestimation is consistently the density term in Eq.~\ref{mHE formula}. In the early stages of the merger, during the first infall of the substructure, the interaction flattens the gas density profile and distorts the overall distribution, effectively reducing the hydrostatic mass estimate. As the system approaches the pericentric passage, the bow shock detaches from the substructure and develops into a runaway merger shock propagating outwards. This outward-moving shock steepens the temperature gradient as it travels through the ICM, leading to a transient overestimation of the hydrostatic mass and hence a positive bias. Therefore, while the temperature gradient steepening contributes to the temporary enhancement of the bias around pericentre, its role remains secondary to the large-scale deformation and compression of the gas distribution during the earlier infall phase. The displacement and reshaping of the gas density field thus emerge as the primary drivers of the hydrostatic mass evolution throughout the event. This supports the interpretation that the characteristic mass bias evolution associated with major mergers, as reported in Sect.~\ref{bias trend}, is primarily a morphological and dynamical effect induced by substructure-driven asymmetries and transient configurations, rather than a consequence of purely thermodynamical processes. The predominant role of the density in shaping the hydrostatic mass estimate is a qualitative feature consistently observed across all merger events in the sample presented in Fig.~\ref{merger bias trend}.

As the merger remnant undergoes relaxation over the subsequent several gigayears, the hydrostatic mass bias progressively approaches its pre-merger levels. Nonetheless, even at late times, the bias remains notably negative, indicating the persistence of a long-lived non-thermal pressure support component. This residual bias is likely sustained by ongoing turbulence and bulk motions within the ICM, which continue to violate the assumptions of HE, as previously suggested by \citet{Lau_2009} and \citet{Nelson_2012} (see also, e.g., \citealp{Valles-Perez_2021} for the evolution of solenoidal turbulence during mergers). This persistence is further illustrated in the $z=0$ panels of Fig.~\ref{fig:multirow_merger}, where despite the morphologically relaxed state of the system, presents a strong ($b_{R_{\mathrm{vir}}}\sim -0.35$) mass bias. While within the inner regions ($r < 0.5R_{\mathrm{vir}}$) the hydrostatic mass bias remains below $\sim11\%$, the outer regions exhibit a much larger discrepancy between the true and the hydrostatic mass estimates. This radial trend, reported in Sect. \ref{merger trend fit}, reflects the delayed equilibration in the cluster outskirts, where residual merger-induced motions and structural asymmetries continue to hinder the system’s convergence towards HE.


\section{Conclusions and discussion}\label{conclusions}

Using a high-resolution cosmological hydrodynamical simulation, we have investigated the connection between the dynamical state of galaxy clusters and the hydrostatic mass bias at  $z=0$, as well as its time evolution during cluster major mergers. By selecting a 
sample of 74 isolated major merger events occurring between $z=1.5$ and $z=0$, we have analysed the physical mechanisms driving the transient and long-term behaviour of the mass bias. Our key findings can be summarised as follows:

\begin{itemize}

    \item At $z=0$, we observe a mild trend for more dynamically relaxed galaxy clusters, which tend to exhibit smaller (but still considerably negative) hydrostatic mass biases at the virial radius. However, the large dispersion and the difference of evolutionary stages between the sampled clusters prevents to draw a strong correlation.
    
    \item Over  clusters lifetime, the hydrostatic mass bias exhibits highly non-monotonic behaviour, generally reflecting the stochastic nature of mass accretion. Its evolution is tightly linked to the cluster mass growth history, where abrupt increases of the mass are often accompanied by dips and spikes in the bias, caused by events that also trigger dynamical perturbations.

    \item During major mergers, the hydrostatic mass bias follows a remarkably consistent trend. As the secondary halo approaches and enters the host volume, the bias deepens, reaching an average minimum of \( b_{R_{\mathrm{vir}}} \sim -0.3 \) evaluating the virial volume. This is followed by a rise, often overshooting zero, reaching positive values in roughly half of the events, and then gradually stabilising near \( b_{R_{\mathrm{vir}}} \sim -0.2 \) after $1.5 \tau_\mathrm{vir}$ \cite[see also ][]{Nelson_2012,Bennett_2022}.

    \item Extending the analysis to different radii within the virial volume, we find that the bias trend during mergers is present from $0.25R_{\mathrm{vir}}$ out to $R_{\mathrm{vir}}$, albeit with varying amplitudes and timescales. In particular, by fitting the central oscillation with a sinusoidal model over a time interval corresponding to one period, we find systematic trends in the fit parameters: the amplitude increases with radius, while the frequency decreases, and the baseline becomes progressively more negative toward larger radii. This behaviour indicates that the impact of major mergers on the hydrostatic mass bias becomes progressively more pronounced when moving from the cluster core toward the virial radius.
    
    \item Despite the expected relation between the impact parameter, the mass ratio and the evolution of bias across mergers drawn from our toy-model, the same trends are not present in our data, probably as a consequence of additional assembly processes, more continuous filamentary accretion, etc.

    \item The physical origin of the observed mass bias evolution can be traced to the structural reconfiguration of the ICM. As the merging halo penetrates the host, the density and pressure profiles flatten in the outskirts, leading to a mass underestimate. Subsequently, the outward-propagating merger shock steepens both the density and temperature gradients, producing a temporary and localised hydrostatic mass overestimation. Consistently across all events studied, the evolution of the mass bias is primarily driven by geometric and dynamical distortions in the gas density distribution rather than by purely thermodynamical processes.

    \item This implies that the characteristic fluctuation of the hydrostatic mass is largely governed by the transient morphological disturbances of the ICM, relegating the modifications of its temperature structure to a more secondary role. This underscores the importance of accounting for these morphological effects when observationally estimating masses based on X-ray or SZ data, as the (relatively customary) determination of the merging stage of a cluster may offer a way to importantly eliminate residual biases that cannot be captured by a universal bias value.

    \item Even several Gyr after the merger, and despite an overall relaxation of the system, a residual negative hydrostatic mass bias persists at all radii, albeit becoming more significant as the radius increases. This suggests that the recovery of hydrostatic equilibrium is a gradual, radially-dependent process, and that non-thermal pressure support and morphological asymmetries may remain significant in cluster outskirts.

    \item Although we find a weak correlation between hydrostatic mass bias and the dynamical state, our results give a robust estimation of bias values at different radii and phases of cluster mergers, namely, merging, just after merging and pre and post-merger. Besides, we propose a basic sinusoidal fit, or an alternative step-wise fit for the premerger, merging and postmerger states, for the mass bias as a function of time that could be used as a proxy to estimate the hydrostatic mass bias from cluster observational data.

\end{itemize}

These findings reinforce the interpretation that hydrostatic mass bias in galaxy clusters, particularly during and after mergers, is governed not only by thermalisation processes, but also by morphologic and kinematic disturbances of the ICM. They further emphasise the limitations of hydrostatic equilibrium assumptions in precision cosmology using cluster mass estimates.

When interpreting our results, it is important to take into consideration  some of the caveats of our simulations and analysis. Notably, the simulation does not include feedback from AGN  nor star formation processes, which are known to influence the thermodynamical state of the ICM. However, we have mainly focused on $r \in (0.25 R_\mathrm{vir}, R_\mathrm{vir})$, away from the region most impacted by these processes. Future work should aim to account for non-thermal pressure support and magnetic field effects, as well as quantify filamentary connectivity, all of which may influence the relaxation state and mass accretion histories of galaxy clusters.

\begin{acknowledgements}

      This work has been supported by the Agencia Estatal de Investigación Española (AEI; grant PID2022-138855NB-C33), by the Ministerio de Ciencia e Innovación (MCIN) within the Plan de Recuperación, Transformación y Resiliencia del Gobierno de España through the project ASFAE/2022/001, with funding from European Union NextGenerationEU (PRTR-C17.I1), and by the Generalitat Valenciana (grant PROMETEO CIPROM/2022/49).
     IBL acknowledges financial support from the Generalitat Valenciana (grant CIACIF/2023/466) and DV acknowledge support from Universitat de València through  Atracció de Talent fellowship. Simulations have been carried out using the supercomputer Lluís Vives at the Servei d'Informàtica of the Universitat de València. 
     
\end{acknowledgements}


\bibliographystyle{aa}
\bibliography{bibliography}

\appendix

\section{How overlapping major mergers affect the hydrostatic mass bias}
\label{appendix overlaping mergers}

In this Appendix, we present the case study of a dynamically complex galaxy cluster, \texttt{CL4}, which was not within the former studied merger sample due to its mergers being non-isolated and close to simulation cut-off boundaries. Despite this, its intricate history of consecutive and overlapping mergers provides valuable insight into the non-linear and cumulative impact of merger dynamics on the hydrostatic mass bias and accretion history.

\begin{figure}[h]
    \centering
    \includegraphics[width=\linewidth]{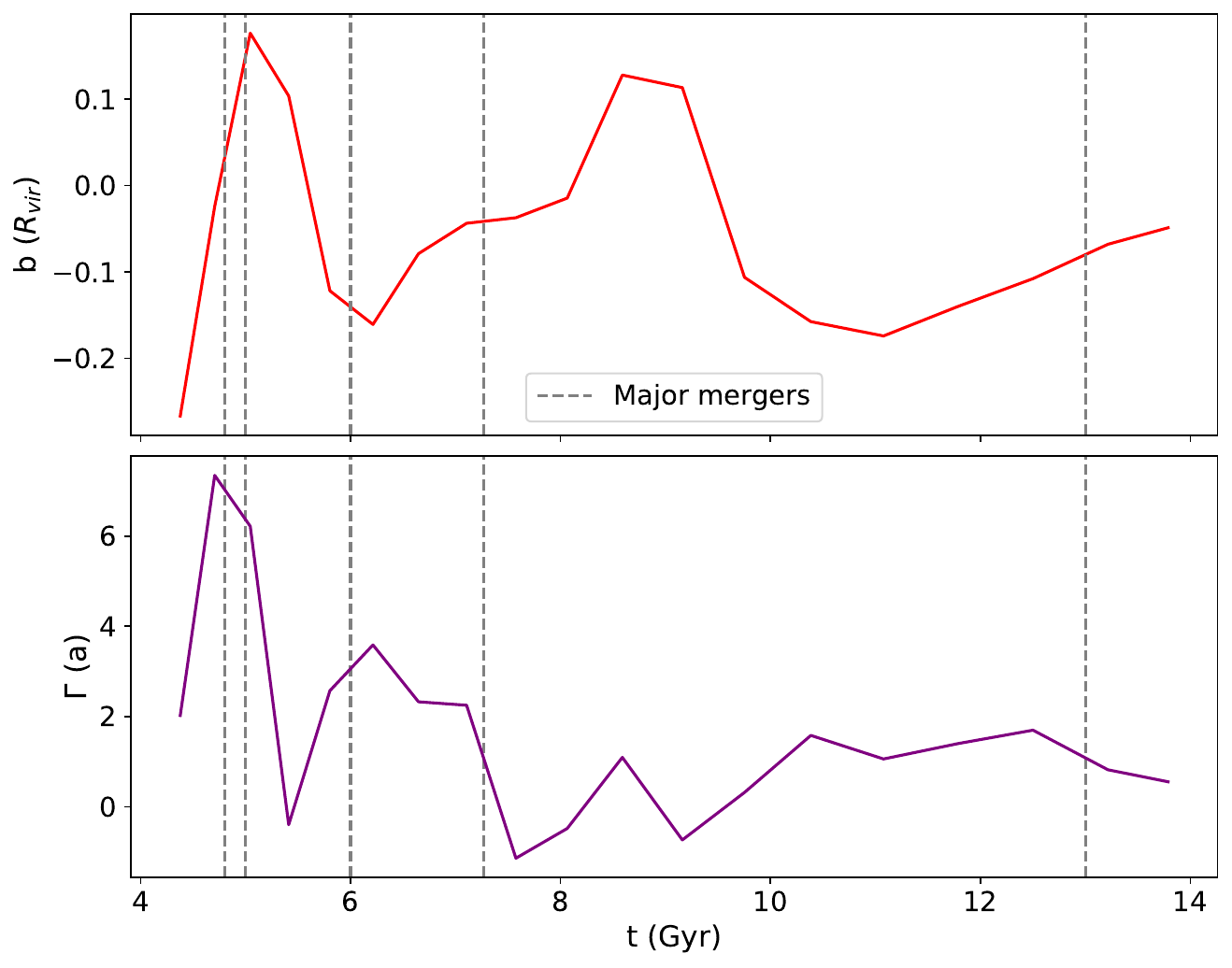}
    \caption{Time evolution of the galaxy cluster \texttt{CL4} with multiple overlapping non-isolated major mergers. {\it Top panel}: Evolution of the hydrostatic mass bias evaluated at $R_{\mathrm{vir}}$. {\it Bottom panel:} $\Gamma(a)$ evolution over the virial volume. The time coordinate is the cosmic time of the simulation. The vertical dashed lines correspond to the major merger events.}
    \label{fig:gc121_bias_gamma}
\end{figure}

Figure~\ref{fig:gc121_bias_gamma} shows the time evolution of the hydrostatic mass bias $b(R_{\mathrm{vir}})$ (top panel) and the logarithmic mass accretion rate $\Gamma(a)$ (bottom panel) from redshift $z \sim 1.5$ to $z = 0$. The vertical dashed lines mark the identified major merger events.

At early cosmic times ($t \sim 4.5$ Gyr), \texttt{CL4} undergoes a pronounced multiple merger episode, with two infalling halos of significant mass ratios ($q \sim 0.8$ and $q \sim 0.9$). This violent interaction is accompanied by the subsequent sharp peak in both the hydrostatic mass bias and $\Gamma(a)$, indicating a strong coupling between mass accretion and hydrostatic disequilibrium. It is worth noting that the merger time inferred from the simulation may be slightly delayed due to limitations in time resolution and cut-off boundaries. The actual start of the interaction could precede our redshift interval.

Following the initial peak associated with the outward propagation of merger shocks and partial ICM thermalisation, the mass bias becomes more negative rapidly, likely due to the flattening of the external regions of the host halo produced by the infalling halo corresponding to the merger located around $t \sim 6$ Gyr. Then, while the bias rises, tentatively following the trend proposed in Sect. \ref{bias in merger}, a new major merger takes place at $t \sim 7.1$ Gyr, with a mass ratio $ \sim 0.3$. Although close to the  threshold  used to separate major and minor mergers, this event has a non-negligible dynamical impact. Instead of a sharp dip, the bias curve exhibits a plateau extending almost a  Gyr, suggesting a more appeased ICM response. This behaviour is likely due to the overlapping dynamical effects of the two mergers. The second merger begins before the system has fully relaxed from the first, leading to a superposition of responses. This plateau phase highlights how the dynamical memory of earlier mergers can modulate the response to subsequent events, making the interpretation of individual bias features more complex.

The corresponding $\Gamma(a)$ evolution further supports this picture. The mass accretion rate remains elevated during this entire interval, reinforcing the idea that \texttt{CL4} is undergoing sustained accretion, which prevents a quick return to HE, and showing that events are overlapping, leaving no respite for the cluster.

At $t \sim 9$ Gyr, the bias curve peaks again. This coincides with a decrement in $\Gamma(a)$, suggesting the cluster is entering a post-merger relaxation phase. The positive excursion in bias is likely due the phenomena explained in Sect. \ref{bias trend causes}, which steepens the pressure and temperature gradients and temporarily increases the hydrostatic mass estimate, a phenomenon already detected in simulations by, for instance, \citet{Nelson_2012} and \citet{Biffi_2016}.

After this phase, the bias curve resumes its decline, stabilising at a moderately negative value ($b\sim-0.1$ to $-0.2$) at late times ($t > 12$ Gyr). This residual bias supports the notion of long-lived non-thermal pressure components and asymmetries in the ICM. Despite the moderation in accretion activity, the system does not return to full HE, likely due to residual bulk motions and incomplete thermalisation in the outer regions (just as the sample depicted in \ref{bias in merger}).

These findings reaffirm  important aspects regarding the evolution of the hydrostatic mass bias. First, the presence of a plateau in the bias curve demonstrates how consecutive mergers can interact non-linearly, producing a cumulative response in the ICM that smooths out individual features typically seen in isolated events. This ``merger stacking'' effect suggests that the ICM retains a memory of recent disturbances, making it difficult to disentangle the impact of each merger and justifying our choice to select isolated events in order to see the cleanest possible trend (within the intrinsic complexity of the mergers we have already discussed). 

Finally, even several Gyrs after the last major interaction, \texttt{CL4} still exhibits a significant residual bias, particularly in the outskirts, as do the subjects in the statistical sample studied, underscoring the long timescales over which the ICM relaxes and highlighting the limitations of assuming HE in recently merged or dynamically active clusters. Taken together, the case of \texttt{CL4} illustrates, as a qualitative representation of all discarded clusters with non-trivial merger situations, how the bias is shaped by a complex interplay of mass accretion history, merger timing, radial dependence and concatenation of events.

\end{document}